\documentclass{iopjournal}
\usepackage{amsmath}
\usepackage{siunitx}
\usepackage{float} 
\usepackage{cleveref}
\crefname{appendix}{appendix}{appendices}
\Crefname{appendix}{Appendix}{Appendices}

\usepackage[backend=bibtex, style=phys]{biblatex}
\begin{document}

\articletype{Paper} %	 e.g. Paper, Letter, Topical Review...

\title{Feasibility of a Flexible, Hybrid Tokamak-Stellarator Experiment using an Axisymmetric Dipole Coil Array}

\author{Jacob Halpern$^{1,*}$\orcid{0000-0001-9370-8160}, Mohammed Haque$^1$\orcid{0009-0007-6194-1911}, Elizabeth Paul$^{1}$\orcid{0000-0002-9355-5595}, Carlos Paz-Soldan$^1$\orcid{0000-0001-5069-4934}, Rithik Banerjee$^1$\orcid{0009-0009-8395-3467}, Talia Angles$^1$\orcid{0009-0006-0013-1889}, Frederick Sheehan$^1$\orcid{0009-0001-6094-3639}, Ian Stewart$^{1}$\orcid{0000-0001-5871-4258}, and Alan Kaptanoglu$^{2,3}$\orcid{0000-0002-6337-2907}}

\affil{$^1$Columbia University, New York NY, USA}
\affil{$^2$New York University, New York NY, USA}
\affil{$^3$University of Washington, Seattle WA, USA}

\affil{$^*$Author to whom any correspondence should be addressed.}

\email{jmh2363@columbia.edu}

\keywords{hybrid tokamak-stellarator, dipole coils, quasi-axisymmetry, stellarator optimization}\newline

\begin{abstract}
We demonstrate the design of a flexible, university-scale hybrid tokamak-stellarator experiment based on an axisymmetric array of planar HTS dipole coils. Because the coil array has few geometric degrees of freedom, we use single-stage optimization of the coil currents initialized from two-stage solutions to obtain mutually consistent equilibria and coil sets within estimated engineering limits. We find that the field error sets a minimum coil-plasma distance which bounds the axisymmetric envelope available to the boundary, within which rotational transform, volume, coil current, and quasi-symmetry (QS) error trade off against one another. Tighter current limits delocalize the non-axisymmetric shaping and raise QS error at fixed transform. From this single coil array we obtain a broad range of equilibria---quasi-axisymmetric vacuum stellarators with edge $\iota$ up to 0.2, finite-$\beta$ hybrids with realistic profiles reaching on-axis $\iota\approx1$ and vacuum transform relevant for MHD stabilization, and strongly shaped tokamaks with elongation $\kappa\approx1.7$ and triangularity $\delta\approx\pm0.6$---all at peak pointwise coil forces within estimated HTS tolerances. We show the same array can additionally correct TF coil ripple, reducing the number of TF coils required compared to the equivalent tokamak. These results establish the design as a promising platform for hybrid tokamak-stellarator research.
\end{abstract}

\section{Introduction}
\label{sec:introduction}

Tokamaks are the most advanced magnetic fusion concept due to their simple axisymmetric geometry and favorable confinement properties \cite{wurzelContinuingProgressFusion2025}; however, reliance on a toroidal plasma current to generate rotational transform leads to current-driven instabilities, disruptions, and an inherently pulsed device \cite{helanderStellaratorTokamakPlasmas2012, bandyopadhyayMHDDisruptionsControl2025}. Stellarators instead generate rotational transform through 3D shaping of the magnetic field without a plasma current, but this introduces its own challenges: good confinement requires careful optimization over a large parameter space \cite{imbert-gerardIntroductionStellaratorsMagnetic2024}, and the complex coils required to produce the 3D field have historically led to significant cost overruns and delays \cite{chrzanowski2009, boschEngineeringChallengesW7X2018a}. As a result, stellarators remain less developed than tokamaks theoretically and experimentally \cite{wurzelContinuingProgressFusion2025}.

Two complementary approaches address these limitations. The first is to simplify coil geometry: stellarators with some or all planar coils reduce manufacturing complexity \cite{morozLowaspectratioStellaratorsPlanar1997a,pedersenExperimentalDemonstrationCompact2006,clarkProtoCIRCUSTiltedcoilTokamak2014,suzukiDesignSimpleStellarator2021a}, including recent designs featuring non-encircling dipole coils optimized for good confinement \cite{gatesStellaratorFusionSystems2025,wuPlanarCoilOptimization2025,krugerCoilOptimizationMethods2025,kaptanogluReactorscaleStellaratorsForce2025,baillodEnhancingStellaratorAccessibility2025a}. The second is hybrid tokamak-stellarator operation, which uses external rotational transform to suppress current-driven or tearing instabilities \cite{teamStabilization211980,fuVerticalStabilityCurrentcarrying2000,archmillerSuppressionVerticalInstability2014,pandyaLowEdgeSafety2015,meng2025a,li2026}, eliminating a central disadvantage of pure tokamak operation. Since coil complexity scales with the vacuum rotational transform \cite{kuNonaxisymmetricShapingTokamaks2009}, producing only a fraction of the total transform with coils also eases engineering requirements; however, the 3D component must be optimized for quasi-symmetry or omnigenity to avoid the orbit losses that unoptimized 3D shaping would otherwise introduce. Several hybrid experiments have been proposed or built \cite{hartwellDesignConstructionOperation2017a, yamazakiTOKASTARTokamakstellaratorHybrid1985,oishiConstructionPlasmaConfinement2014,liDesignExternalRotational2024, hennebergCompactStellaratortokamakHybrid2024a,hennebergVarietyCoilSets2025,liangDesign3DEquilibria2025} and can achieve lower aspect ratios than traditional stellarators \cite{hennebergCompactStellaratortokamakHybrid2024a}.

In this study, we demonstrate the feasibility of a hybrid tokamak-stellarator experiment consisting of axisymmetric TF coils and a dipole coil array, along with traditional poloidal field coils and a central solenoid. We base our analysis on the geometry of the HBT-EP experiment at Columbia University \cite{maurerHighBetaTokamakextended2011}, using its $\qty{1.0}{\meter}$ major radius and vacuum vessel as reference dimensions and targeting stellarator plasmas with similar aspect ratio of $\approx 7$. Our physics modeling assumes high-temperature superconducting (HTS) tape constraints, though the coil optimization framework is conductor-agnostic and the results scale to other current-carrying technologies.

The primary advantage of the design is that the dipole coils are mounted on an axisymmetric vessel, imposing a near-axisymmetric structure on the coil array. The only departure from exact axisymmetry arises from the discrete toroidal gaps between coils, which produce benign high toroidal mode number field fluctuations analogous to TF coil ripple in existing tokamaks. Because it is not tied to a single equilibrium, the array can generate many stellarator and tokamak configurations, and the dipole coil currents can be varied to create optimized axisymmetric and non-axisymmetric configurations with different physics properties such as different MHD stability, turbulent transport, and divertor structures. Furthermore, this configuration allows the toroidal rotation of 3D equilibria past a fixed diagnostic, enabling fully 3D diagnosis across multiple shots. With a central solenoid, the device can also study how bootstrap current modifies optimized stellarator equilibria---directly relevant to reactor scenarios---and how applied 3D fields can stabilize MHD modes in tokamak operation. Although less efficient than equilibrium-specific 3D coils, the axisymmetric array trades this efficiency for flexibility, making it attractive as an academic platform rather than a reactor-relevant design. This design can be seen as a middle-ground between other ``pixelated'' magnetic field approaches such as traditional tokamak saddle coils used for error field correction/resonant magnetic perturbations \cite{yangTailoringTokamakError2024} and permanent magnet stellarators \cite{qianSimplerOptimizedStellarators2022,qianDesignConstructionMUSE2023}.

This paper demonstrates the flexibility of the design configuration \cite{leeStellaratorCoilOptimization2022, parraFlexibleStellaratorPhysics2024,yamaguchiInnovativeStellaratorVariable,yuProgrammableStellaratortokamakHybrid2026}, which spans quasi-axisymmetric vacuum equilibria with edge rotational transform up to $\iota \approx 0.2$ and finite-$\beta$ configurations with realistic plasma profiles while adhering to approximated engineering limits based on the pointwise coil forces. Because the constrained coil geometry makes traditional two-stage optimization insufficient, we use two-stage results only as hot-start initial conditions for a single-stage algorithm that simultaneously optimizes the plasma surface and coil currents. We explore tradeoffs between rotational transform, QS error, volume, and coil-current constraints, finding that field-error and coil-current thresholds jointly constrain the minimum and maximum coil-plasma distance, bounding the axisymmetric envelope available to the surface, with rotational transform and volume competing for it. In \cref{sec:design-specification}, we describe the coil geometry and engineering constraints. In \cref{sec:stellarator-optimization}, we present our optimization methods and results, including design tradeoffs and both vacuum and finite-$\beta$ equilibria. In \cref{sec:tokamak-benefits}, we show that the same coil array can correct TF coil ripple and create shaped tokamak equilibria, demonstrating the platform's versatility beyond stellarator operation.

\section{Design Concept}
\label{sec:design-specification}

\begin{figure}[H]
 \centering
    \includegraphics[width=\textwidth]{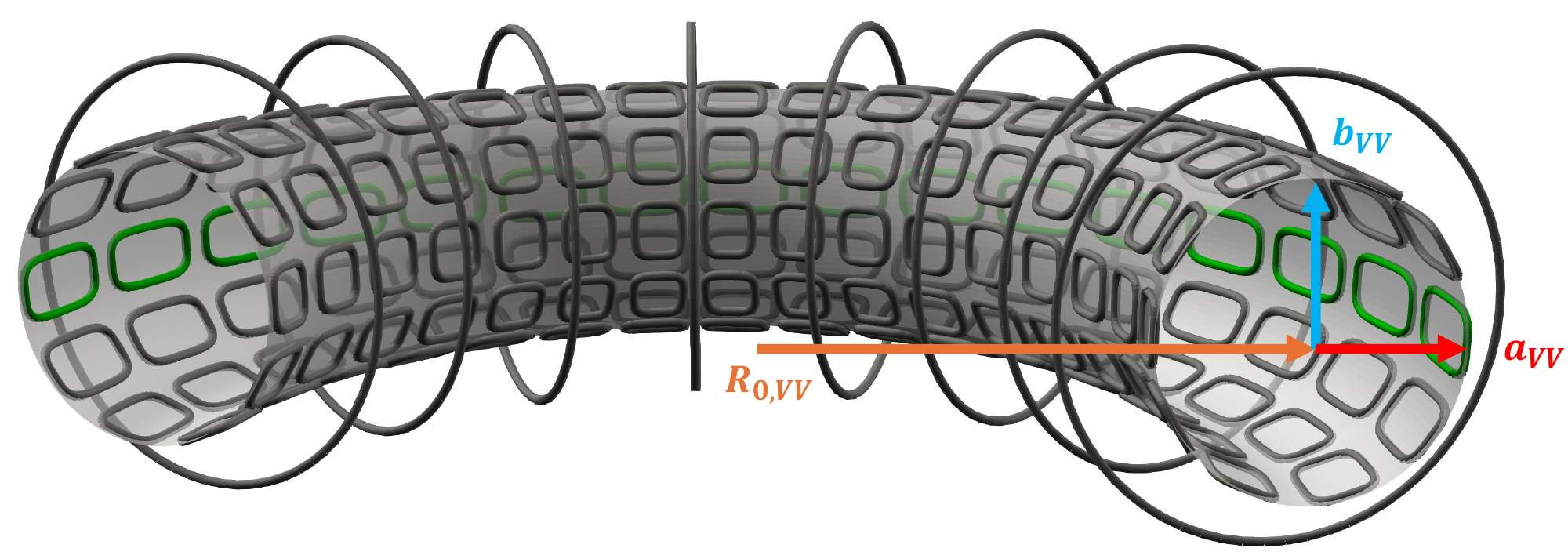}
\caption{Illustration of the design over a $\qty{180}{\degree}$ toroidal sector showing the elliptical vacuum vessel and coils in light and dark grey, respectively. We have labeled the free vacuum vessel geometry parameters, $R_{0,VV}$, $a_{VV}$, and $b_{VV}$. The outboard coil array we remove to demonstrate sparsity in \cref{subsec:sparsity} is shown in green. We do not show any central solenoid or vertical field coils, which are not considered in the vacuum stellarator optimization.}
 \label{fig:coil_geometry}
\end{figure}

In this section, we describe the coil geometry and engineering constraints. We show a schematic of an example coil set in \cref{fig:coil_geometry}.

The dipole coil arrays are initialized on an axisymmetric vacuum vessel with major radius $R_{0,VV}$ and an elliptical cross section of major and minor axes $a_{VV}$ and $b_{VV}$, respectively, and are labeled in \cref{fig:coil_geometry}. As HBT-EP's vacuum vessel has a major radius of $\qty{1.0}{\meter}$ and minor radius of $\qty{0.25}{\meter}$, we enforce that the vacuum vessel is approximately the same size, i.e. $R_{0,VV} - a_{VV} > \qty{0.7}{\meter}$ and $R_{0,VV} + a_{VV} < \qty{1.3}{\meter}$. Within these bounds, we treat the vacuum-vessel geometry as free rather than fixed to the existing HBT-EP dimensions, as replacing the existing vacuum vessel is relatively inexpensive compared to the cost of the HTS material for the coils. We explore this freedom through random sampling in \cref{subsec:stage-II}.

We similarly use an elliptical representation for the toroidal field coils with major and minor axes $a_{TF}$ and $b_{TF}$, respectively. We initialize $N_{TF}$ coils equally spaced in the toroidal angle. We fix a constant spacing of $\qty{15}{\centi\meter}$ between the vacuum vessel and TF coils, i.e. $a_{TF} = a_{VV} + \qty{15}{\centi\meter}$ and $b_{TF} = b_{VV} + \qty{15}{\centi\meter}$, matching the current HBT-EP specifications. For this study, we neglect the fields from the central solenoid and axisymmetric poloidal-field coils except through the induced toroidal plasma current in \cref{subsec:finite-beta-optimization} and \cref{subsec:shaped-tokamak-equilibria}. We fix the TF coil currents to produce a toroidal field of $\qty{0.5}{\tesla}$ on axis, which is a reasonable field strength for HTS coils and is larger than the field strength of HBT-EP; however, our results can be generalized to other field strengths as the required magnetic field from the dipole coils scales linearly with the coil currents.

We initialize an array of $N_{\theta}$ coils in the poloidal direction and $N_{\phi}$ coils in the toroidal direction. The coils are described as rounded rectangles \cite{nashPrototypingTestCanis2025b} of width $w_{\theta}$ in the poloidal direction and $w_{\phi}$ in the toroidal direction, which are uniquely determined given $R_{0,VV}$, $a_{VV}$, $b_{VV}$, $N_{\theta}$, $N_{\phi}$, and their spacing. The coil centers are equally spaced, lie on the vacuum vessel, and are locally tangent to it. We fix a minimum edge-to-edge separation of $\qty{5}{\centi\meter}$ for the coil filaments to accommodate the finite build of the coil \cite{nashPrototypingTestCanis2025b}, corresponding to a maximum circular cross-sectional radius of $\approx\qty{2.5}{\centi\meter}$. Note that this feature makes $w_{\phi}$ larger on the outboard side than on the inboard, as seen in \cref{fig:coil_geometry}. We reserve a detailed description of how this initialization was implemented in the SIMSOPT code \cite{landremanSIMSOPTFlexibleFramework2021} in \cref{sec:windowpane-array}. We choose $N_{\phi}$ as a multiple of $N_{TF}$ to reduce field errors, as the dipole coils consistently lie between the TF coils and can efficiently heal TF coil ripple for both stellarator and tokamak operation (see \cref{subsec:tf-ripple}); it is also optimal from an engineering perspective as the array could be segmented at the same points relative to the TF coils. We have found $N_{\phi} = 32$ and $N_{TF} = 16$ to be optimal in creating dipole coils with both reasonable size and sufficient degrees of freedom to effectively optimize the magnetic field. This configuration can also support several relevant equilibrium field periodicities, i.e. $1, 2, 4, 8\dots$.

The maximum coil current is our primary electromagnetic engineering constraint, as it directly determines the forces and torques on the coil support structure. We estimate these quantities from our optimized coil sets using an analytic self-force model \cite{hurwitzEfficientCalculationSelfMagnetic2024a, landremanEfficientCalculationSelf2025} in SIMSOPT \cite{landremanSIMSOPTFlexibleFramework2021} with the $\qty{2.5}{\centi\meter}$ circular cross-sectional radius above for the dipole, and $\qty{5}{\centi\meter}$ for the TF coils. HTS tape tolerates maximum pointwise force per unit length $|d\mathbf{F}/dl|$ of $4$–$8\times10^5~\unit{\newton\per\meter}$ \cite{zhaoStructuralModelingREBCO2022a, riva2023, hartwigSPARCToroidalField2024}, and we select current thresholds in post-processing to remain well below this range. We note that $|d\mathbf{F}/dl|$ is used here as a preliminary proxy for coil limits---it is straightforward to evaluate analytically and has established HTS tolerance values in the literature. However, the net forces and torques on the coil support structure are equally relevant engineering constraints and may in practice be the binding limit; these quantities are harder to bound without detailed finite element analysis of the support structure that is beyond the scope of this study. Our current thresholds in \cref{subsec:single-stage-optimization} are therefore chosen conservatively, with the expectation that a full structural analysis will refine these limits in future design iterations.

This approach differs from most stellarator coil-optimization routines where coils have individual geometric degrees of freedom \cite{zhuNewMethodDesign2017a}. This includes other studies including planar non-encircling dipole coils, where each coil's position/orientation is optimized to create the desired field or adhere to engineering constraints \cite{kaptanogluReactorscaleStellaratorsForce2025,swanson2026}. In our case, the coil geometry is varied only through the three continuous vacuum vessel geometric parameters and the discrete number of poloidal coils $N_\theta$, with all other quantities above fixed or dependent on one or more free parameters. We will find that this limited parameter space for the coil geometry prevents the use of traditional optimization techniques where the coil and plasma are optimized separately and motivates the use of a simultaneous approach where the dipole current degrees of freedom are incorporated into the equilibrium optimization.

\section{Stellarator Optimization}
\label{sec:stellarator-optimization}

Many stellarator optimization methods use a two-stage approach: first the plasma equilibrium is optimized in stage-I for desirable physics properties \cite{landremanMagneticFieldsPrecise2022} and then the coil set is optimized in stage-II to create this equilibrium magnetic field subject to engineering constraints \cite{merkelSolutionStellaratorBoundary1987,stricklerDesigningCoilsCompact2002,landremanImprovedCurrentPotential2017b, zhuNewMethodDesign2017a, lobsienStellaratorCoilOptimization2018, fuGlobalStellaratorCoil2025}. This decoupling can make each individual step more tractable, but it does not guarantee a coil-realizable stage-I equilibrium under realistic engineering limits. Single-stage algorithms instead optimize equilibrium and coils simultaneously \cite{hennebergCombinedPlasmaCoil2021} and have received significant development and usage in recent years \cite{giulianiSinglestageGradientbasedStellarator2022,giulianiDirectStellaratorCoil2023, jorgeSinglestageStellaratorOptimization2023, baillodIntegratingNovelStellarator2025}. Single-stage algorithms are particularly advantageous in this case as engineering constraints from the coil set can be directly incorporated into the equilibrium optimization, but typically are accompanied by an increase in computational cost. For this device, the relevant goals are to maximize vacuum rotational transform and minimize QS error for confinement, while maximizing plasma volume and adhering to reasonable engineering constraints on the coil set---a combination that, as we will show, is only achievable through the single-stage approach.

Constrained coil geometry makes single-stage optimization especially important \cite{baillodIntegratingNovelStellarator2025}. In our design, the coil geometry has only four degrees of freedom---one of which is discrete (the number of poloidal coils)---making it extremely likely that a stage-I equilibrium is incompatible with the coil set. We therefore adopt a multi-stage approach: we first generate hot-start configurations with a mutually consistent equilibrium and coil set, then refine them using single-stage optimization. While single-stage algorithms can suffer from heightened sensitivity to initial conditions, we do not observe this in our analysis. We found that the single-stage optimizer converges to similar solutions from varied initial conditions, which we attribute to the linear relation between dipole coil currents and magnetic field. This contrasts with modular coil optimization, where geometric variation produces a highly non-convex landscape and single-stage methods are considerably more sensitive to initialization \cite{baillodIntegratingNovelStellarator2025}.

\subsection{Generating Initial Conditions}
\label{subsec:initial-conditions}

While we found single-stage optimization is necessary to obtain an optimal solution due to the constrained nature of the coil set, it would be computationally intractable to use it to optimize over all of our degrees of freedom and it generally requires a good initial condition for convergence. As a result, we first use a two-stage approach to identify a reasonable configuration of an equilibrium and coil set geometry. In this section, we describe the two-stage optimization approach we use to generate this initial condition.

\subsubsection{Equilibrium Optimization}
\label{subsec:stage-I}

The goal of our stage-I optimization is to produce an equilibrium that can initialize the single-stage routine---one with sufficient rotational transform to be a useful starting point, but smooth enough that a compatible coil set can be found within engineering limits. These two requirements compete, as near-axis theory shows that rotational transform is generated by a combination of axis torsion (non-planarity of the magnetic axis) and elliptical cross section rotation \cite{imbert-gerardIntroductionStellaratorsMagnetic2024}. Both require increasing the non-axisymmetric deformation of the plasma boundary, directly reducing the volume that can fit within a fixed axisymmetric envelope. This establishes a fundamental tradeoff between vacuum rotational transform and plasma volume, which we visualize in \cref{sec:near-axis-tradeoffs}; similar findings have been observed in other hybrid studies \cite{hennebergCompactStellaratortokamakHybrid2024a}. Increasing rotational transform is also costly for finding a consistent coil set; we found that it manifests as regions of sharp local curvature, which typically require small coils with large current to reproduce, and surfaces with large variation in the axis that causes the equilibrium to approach the coils closely, where strong local field rippling increases stage-II error. 

We optimize for quasi-axisymmetric (QA) equilibria, the most tokamak-like equilibria in both physics and shape \cite{boozerStellaratorsPathITER2008a,plunkQuasiaxisymmetricMagneticFields2018,plunkPerturbingAxisymmetricMagnetic2020, hennebergVarietyCoilSets2025}. We describe the equilibrium as a stellarator symmetric, double Fourier series in cylindrical coordinates $(R, \phi, Z)$ with the cross section in the $(R, Z)$ plane parametrized by the poloidal angle $\theta$,
\begin{eqnarray}
    R(\theta,\phi) = \sum_{m=0}^{m_{max}} \sum_{n=-n_{max}}^{n_{max}} R_{mn} \cos(m\theta - n N_{fp} \phi), \\
    Z(\theta,\phi) = \sum_{m=0}^{m_{max}} \sum_{n=-n_{max}}^{n_{max}} Z_{mn} \sin(m\theta - n N_{fp} \phi),
\end{eqnarray}
where $R_{mn}$ and $Z_{mn}$ are the Fourier coefficients, $m_{max} = 2$ and $n_{max} = 2$ are the maximum poloidal and toroidal harmonics, and $N_{fp} = 2$ is the number of field periods. Note that our methods are valid for any $N_{fp}$ supported by the coil set.
The degrees of freedom are the Fourier coefficients of the equilibrium, $\mathbf{x} = \{R_{mn}, Z_{mn}\}$, and we formulate our optimization as a nonlinear-least-squares problem 
\begin{equation}
    \min_{\mathbf{x}} f_{QS}^2 + w_{\iota} \left( \iota - \iota_T \right)^2 + w_V \left( V - V_T \right)^2 + w_{d} f_{d}^2 + w_{\kappa} f_{\kappa}, \label{eq:stage-I-objective-function}
\end{equation}
where $f_{QS}$ is the sum over discrete flux surfaces $s_i$ for the flux-surface-averaged, two-term, QS objective in vacuum
\begin{equation}
    f_{QS} = \sum_{s_i} \left\langle \frac{1}{B^3} \left(N - \iota M \right) \mathbf{B}\times \nabla B \cdot \nabla \psi - M G \mathbf{B} \cdot \nabla B \right\rangle \label{eq:two-term-qs},
\end{equation}
$\iota$ and $\iota_T$ target the on-axis rotational transform, $f_{d}$ penalizes minimum vacuum vessel to plasma distances $d_{VV-plas,min}$ less than a threshold $d_T$,
\begin{equation}
    f_{d} = \min(0, d_{VV-plas,min} - d_{T})^2,
\end{equation}
and $f_{\kappa}$ smoothly penalizes the large surface curvatures,
\begin{equation}
    f_{\kappa} =
     \int dS \left(\exp\left( \frac{-\kappa_1 + \kappa_{1,T}}{w_1} \right) + \exp\left( \frac{\kappa_2 + \kappa_{2,T}}{w_2} \right) \right), \label{eq:objective-function-curvature}
\end{equation}
where $\kappa_1$ and $\kappa_2$ are the principal curvatures of the plasma boundary. This term smoothly penalizes $\kappa_1$ and $-\kappa_2$ above thresholds $\kappa_{1,T}$ and $\kappa_{2,T}$, with $w_1$ and $w_2$ controlling the relative penalty on each. The distance and curvature penalties serve as proxies for coil engineering constraints during equilibrium optimization, analogous in spirit to the current penalty we use in single-stage optimization but at a fraction of the computational cost. We set $V_T = \qty{0.3}{\cubic\meter}$ to be comparable to HBT-EP and $M=1, N=0$ for quasi-axisymmetry. The weights $w_V$, $w_d$, and $w_\kappa$ are all set large enough to enforce their respective targets. We empirically found $d_T = \qty{8}{\centi\meter}$, $w_1 = 1$, $w_2 = -0.5$, $\kappa_{1,T} = \qty{1}{\meter^{-1}}$, and $\kappa_{2,T} = \qty{6}{\meter^{-1}}$ to allow coil sets with reasonable stage-II field errors in our subsequent analysis. 
We then set $\iota_T$ to be large and $w_\iota$ such that the penalty is several orders of magnitude smaller than the volume, distance, and curvature terms but still an order of magnitude larger than the QS objective. This lets the optimizer increase rotational transform until the other constraints become binding regardless of increases in QS error, a limit we examine further in \cref{subsec:single-stage-optimization}. We optimize using the SIMSOPT code \cite{landremanSIMSOPTFlexibleFramework2021}, computing the vacuum magnetic equilibrium using VMEC \cite{hirshmanSteepestdescentMomentMethod1983}. We initialize the minimization from an axisymmetric, circular cross-section boundary, and fix the major radius Fourier coefficient $R_{00}$ to keep the aspect ratio approximately constant. We use the existing HBT-EP vacuum vessel with $R_{0,VV} = \qty{1}{\meter}, a_{VV} = \qty{0.25}{\meter}$ for the distance constraint.

\begin{figure}[H]
    \centering
           \includegraphics[width=\textwidth]{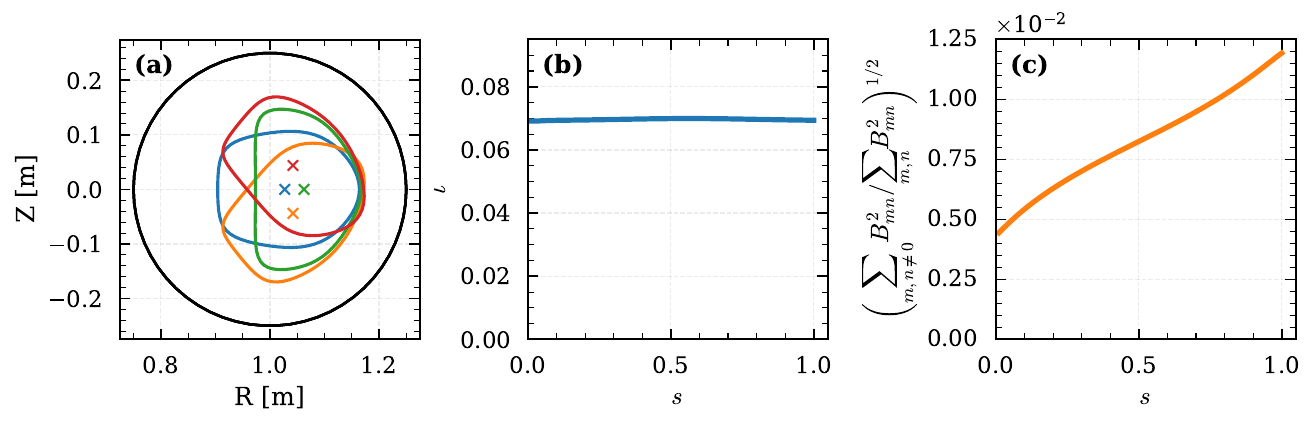}
    \caption{Postprocessing analysis of the equilibrium initial condition obtained from stage-I optimization. (a) Equilibrium cross section at equally spaced toroidal locations, with the magnetic axis location for each toroidal cross section shown as an X of the same color and the vessel constraint in solid black. (b) Rotational transform profile versus normalized toroidal flux (c) Square root of the normalized sum of quasi-axisymmetry-breaking modes in Boozer coordinates versus the normalized toroidal flux \label{fig:stage-I-eq}}
\end{figure}

\Cref{fig:stage-I-eq}(a)-(c) summarizes the optimized stage-I equilibrium, with (a) showing the equilibrium cross section at equally spaced toroidal locations, (b), showing $\iota$ profile, and (c) showing the QS error expressed as the square root sum of QA breaking Fourier modes. We plot this quantity here and throughout the rest of this paper instead of \cref{eq:two-term-qs} because our single-stage optimization in \cref{subsec:single-stage-optimization} employs a different objective, so this more standard metric allows easier comparison of our QS error across our optimizations and other devices \cite{baillod2026}. This equilibrium generates its modest rotational transform of $\iota \approx 0.07$ through a combination of axis torsion and rotating ellipticity. At fixed volume, the axis torsion is limited by the distance penalty, as large variations in the axis cause the boundary to approach the vessel and coils. Similarly, the surface curvature penalty limits the iota generated by rotating ellipticity, as rapid elongated cross-section rotation produces regions of sharp inboard curvature that violate the curvature thresholds. The combined effect of these two constraints, which together encode the engineering accessibility of a consistent coil set, is what causes the optimizer to settle on a modest $\iota$. Despite these constraints, we find in \cref{fig:stage-I-eq}(c) that we can also achieve QS error comparable to other experiments \cite{baillod2026}.

Note that this and all other stage-I optimizations converged to QA equilibria with shaping predominantly localized on the inboard side, consistent with that found in other studies \cite{plunkQuasiaxisymmetricMagneticFields2018,plunkPerturbingAxisymmetricMagnetic2020,hennebergCompactStellaratortokamakHybrid2024a,hennebergVarietyCoilSets2025}. Although QA equilibria with perturbations localized to the outboard side exist, they have been found to have lower QA quality and generate rotational transform less efficiently \cite{liangDesign3DEquilibria2025}, possibly explaining why we did not obtain any in our stage-I optimization. However, we will show that our single-stage optimizations converge to equilibria with shaping in various poloidal locations, demonstrating that this approach is insensitive to the choice of initial condition.

\subsubsection{Coil Optimization}
\label{subsec:stage-II}

We next construct a coil set for the optimized stage-I equilibrium in \cref{fig:stage-I-eq}. To make single-stage optimization tractable, we fix the coil geometry and vary only the currents to optimize the plasma parameters. We therefore use the stage-II approach below to obtain both the coil geometry and initial currents.

We scan coil geometry by randomly sampling the vacuum vessel parameters (major radius, major/minor axis lengths) within physically realistic ranges at fixed $N_\theta$ and initializing the coil set from each sample. With only three continuous geometric degrees of freedom, a modest random sample suffices for the present study. Gradient-based optimization of the vessel geometry, which would require differentiating the field error through the mapping that determines each dipole coil's position and orientation, is left for future work. For each sample geometry, we then perform an optimization of the dipole coil currents to minimize the magnetic field error,
\begin{equation}
    \min_{\mathbf{I}} \int_S \left| \left(\mathbf{B}(\mathbf{I}) + \mathbf{B}_{TF}\right) \cdot \hat{\mathbf{n}} \right|^2 dS,
    \label{eq:stage-II-objective-function}
\end{equation}
where $\mathbf{I}$ is a vector of the dipole coil currents, $\mathbf{B}_{TF}$ is the magnetic field from the fixed TF coils, $\hat{\mathbf{n}}$ is the normal vector to the plasma surface $S$, and the integral is taken over the plasma surface. We omit current regularization at this stage deliberately: adding a current penalty raises the achievable field error floor, which is important for initializing the single-stage algorithm in \cref{subsec:single-stage-optimization}. We verified that the unregularized least-squares problem is well-conditioned for the coil geometries considered here, with no evidence of the degenerate current distributions that would motivate regularization. Current limits are instead enforced implicitly through the Pareto front selection described below.

\begin{figure}[H]
    \centering
    \includegraphics[width=\linewidth]{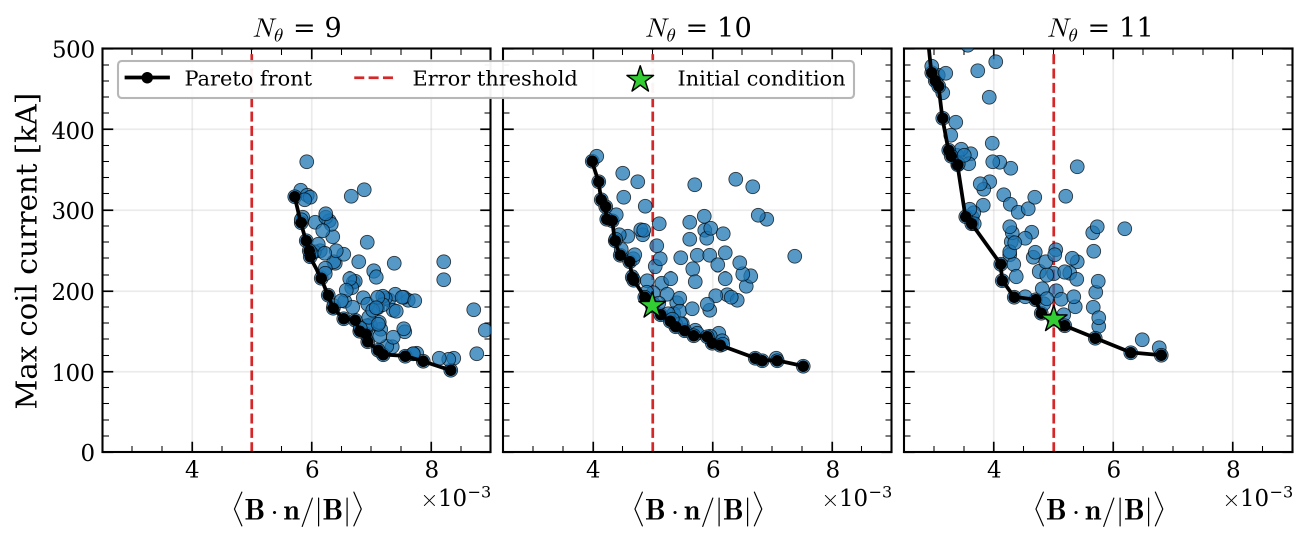}
    \caption{Maximum dipole coil current versus surface average field error for $100$ randomly sampled coil sets with optimized currents for $N_{\theta} = 9, 10, 11$. We plot the Pareto optimal solutions in solid black. The field error threshold required to initialize the single-stage algorithm is plotted as a vertical dashed red line. The minimum current, Pareto optimal solution for each $N_{\theta}$ below this threshold is shown as a green star; one does not exist for $N_{\theta}=9$.}
    \label{fig:stage-II-pareto-front}
\end{figure}

We choose sampling ranges by first computing the axisymmetric envelope of the equilibrium, described by an elliptical cross section with parameters $R_{0,EQ}$, $a_{EQ}$, and $b_{EQ}$, and then sampling from empirically determined ranges of $R_{0,VV} - R_{0,EQ}\in  [-\qty{3}{\centi\meter}, \qty{3}{\centi\meter}]$, $a_{VV} - a_{EQ}\in [\qty{8}{\centi\meter}, \qty{16}{\centi\meter}]$ and $b_{VV} - b_{EQ}\in [\qty{8}{\centi\meter}, \qty{16}{\centi\meter}]$, while filtering out any samples which violate our vessel limits, $R_{0,VV} - R_{0,EQ} < \qty{0.7}{\meter}$ or $R_{0,VV} + R_{0,EQ} > \qty{1.3}{\meter}$. We perform 100 vacuum vessel geometry samples per $N_{\theta}$, plotting the results of this analysis in terms of field error and the maximum coil current for $N_{\theta} = 9, 10, 11$ shown in \cref{fig:stage-II-pareto-front}. We plot the Pareto optimal solution as solid black points.

We require a sufficiently small surface field error for the single-stage algorithm (described in \cref{subsec:single-stage-optimization}) to converge from the stage-I equilibrium; empirically, this threshold is $\approx 0.5\%$, shown as the vertical dashed red line in \cref{fig:stage-II-pareto-front}. Among Pareto-optimal coil sets below this line, lower field error comes at the cost of higher maximum coil current. Since we impose a maximum coil-current limit in single-stage optimization, we initialize from the lowest-current Pareto point that still meets the field-error threshold, rather than pushing stage-II accuracy further. In practice, this choice yielded robust single-stage convergence. These initial conditions are shown as green stars.

\Cref{fig:stage-II-pareto-front} reveals several trends. First, the results emphasize the need for single-stage optimization: even with maximum coil currents exceeding $\qty{500}{\kilo\ampere}$, roughly twice the upper bound estimated in \cref{subsec:single-stage-optimization}, no configurations achieve the $\approx 0.1\%$ field error typically required for two-stage convergence alone \cite{wechsungPreciseStellaratorQuasisymmetry2022, wiedmanCoilOptimizationQuasihelically2024}. This indicates that the equilibrium itself must be co-optimized with the coil set, rather than treated as a fixed target (a trend we find consistent across several stage-I equilibria not shown here).

Second, for fixed $N_{\theta}$, there is a tradeoff between field error and the maximum coil current. Dipole coils placed close to the plasma require smaller currents, but each coil's field is concentrated near its poloidal location, introducing poloidal ripple at the plasma boundary that degrades the achievable field accuracy. When coils are placed farther from the boundary, their fields spread poloidally before reaching the plasma, reducing the per-coil ripple and allowing adjacent coils to cooperatively shape the boundary with greater fidelity. However, because shaping fields associated with poloidal mode number $m$ decay as $r^m$ with distance from the coil \cite{landremanEfficientMagneticFields2016}, producing a given boundary shape from a more distant coil set requires disproportionately larger currents---particularly for highly shaped geometries where high-$m$ modes dominate. We note that this ripple is computed using a filamentary coil model; a finite build coil representation would distribute each coil's field more smoothly and may reduce this effect, loosening the minimum coil-plasma distance constraint.

Finally, we see that as $N_{\theta}$ increases, the Pareto front shifts to lower field error and higher maximum current. This occurs because larger $N_{\theta}$ provides more current degrees of freedom to create the desired equilibrium field, allowing for more accurate fields, but these smaller coils have fields that decay more rapidly with distance from the coil, requiring larger currents to create the same magnetic fields at the equilibrium surface. Notably, the Pareto front shifts sufficiently far to higher field errors for $N_{\theta}=9$ that no Pareto optimal solutions are below the field error threshold, and no single-stage initial conditions could be obtained using this method.

\subsection{Single-Stage Optimization}
\label{subsec:single-stage-optimization}

We fix the coil geometry selected from the stage-II Pareto analysis of \cref{fig:stage-II-pareto-front}, so that all optimized configurations are realizable by the same coil set. Using the single-stage approach, we now optimize both the equilibrium and coil set simultaneously, allowing us to obtain mutually consistent configurations and analyze the effect of varying the rotational transform, coil current constraints, plasma volume, and sparsity.

\subsubsection{Optimization Formulation}
We optimize the vacuum equilibrium using a single-stage optimization routine \cite{giulianiSinglestageGradientbasedStellarator2022, giulianiDirectStellaratorCoil2023,giuliani2024} that couples the plasma and coil degrees of freedom by solving for the magnetic surface $\mathbf{\Gamma}(\theta,\varphi)$ generated by the coil magnetic field. Here, \(\mathbf{\Gamma}\) gives the position vector of points on the surface in Boozer coordinates \((\theta,\varphi)\). Equivalently, the method minimizes the residual of the magnetic-surface partial differential equation (PDE) at fixed enclosed volume $V(\mathbf{\Gamma})$,
\begin{equation}
    R = \frac{1}{2} \int_{\mathbf{\Gamma}} \left|G \mathbf{B} - B^2 \left( \frac{\partial\mathbf{\Gamma}}{\partial\varphi} + \iota \frac{\partial\mathbf{\Gamma}}{\partial\theta}\right)\right|^2 dS
    \label{eq:boozer-residual}
\end{equation}
subject to the constraint $V(\mathbf{\Gamma}) = V_T$. In the above, $G$ is proportional to the poloidal current outside the surface, $\mathbf{B}$ is the magnetic field generated by the coil set, and $V_T$ is the target volume. \Cref{eq:boozer-residual} is solved in Fourier space up to toroidal and poloidal mode numbers $\mathit{ntor}$ and $\mathit{mpol}$. Further algorithmic details are given in \cite{giulianiDirectStellaratorCoil2023}. This technique allows the coil currents to be optimized simultaneously for both engineering constraints and the physics of the equilibrium they create, while retaining differentiability for rapid and reliable optimization.

For this initial design study, we keep the optimization problem minimal; we target quasi-symmetry and rotational transform for confinement and our engineering constraints via the maximum coil current. From our two-stage analysis, we expect there to be at least a tradeoff between rotational transform and coil current; however, effectively optimizing for all three objectives simultaneously while retaining a small residual can be challenging. In this work, we employ a form of epsilon-constraint optimization to best identify the Pareto fronts among these objectives, which has been shown to be effective for similar problems \cite{bindelUnderstandingTradeoffsStellarator2023}. In this case, we choose to minimize a single parameter, the deviation from quasi-symmetry, with all other objectives constrained to be within some small parameter $\epsilon$ of their target values. Our single-stage optimization problem is then formulated as
\begin{equation}
    \min_{\mathbf{I}} f_{QS} \quad \mathrm{subject\ to}\ f_{R}, f_{C}, f_\iota \leq \boldsymbol{\epsilon},
\end{equation}
where $\mathbf{I}$ is the vector of coil currents, $\boldsymbol{\epsilon}$ is a vector of the values of $\epsilon$ for each target, 
\begin{equation}
    f_{QS} = \int_{\mathbf{\Gamma}}dS B^2_{non-QS} \ / \int_{\mathbf{\Gamma}}dS B^2_{QS} \label{eq:QS-error}
\end{equation}
quantifies QS error as the surface-integrated ratio of the non-quasi-symmetric to quasi-symmetric magnetic field,
\begin{equation}
    f_{R} = \max\left({0, R - R_T}\right)^2
\end{equation}
constrains the residual to be below a threshold $R_T$,
\begin{equation}
    f_{C} = \max\left({0, I^p - I^p_T}\right)^2,
\end{equation}
constrains the $p$-norm of the coil currents $I^p = \left(\sum_i |I_i|^p\right)^{1/p}$ to be below a threshold $I^p_T$, and
\begin{equation}
    f_{\iota} = \left(\iota - \iota_T\right)^2,
\end{equation}
penalizes the deviation from the target iota $\iota_T$. All equilibrium quantities evaluated in the objectives above and quoted in this section are evaluated for the boundary surface $\mathbf{\Gamma}$ identified by minimizing \cref{eq:boozer-residual}. Note the difference between the stage-I QS objective over the entire equilibrium volume, \cref{eq:two-term-qs}, and our single-stage objective for evaluated over the boundary. We use $p=20$ so that $I^p$ approximates the maximum coil current---since $|\mathbf{I}|_p \to \max_i |I_i|$ as $p \to \infty$---while remaining a smooth, differentiable surrogate that is easier to optimize. Previous studies \cite{kaptanogluReactorscaleStellaratorsForce2025} have instead penalized coil forces and torques, as these more directly translate into engineering constraints; however, it was found that this mainly introduced variations in the orientation of the coils, which cannot occur in our fixed-geometry optimization, so the current penalty captures the essential content of those penalties while remaining simple to evaluate.

Rather than directly constraining these objectives, we employ scalarization by using a large weight to enforce the constraints terms, making the optimization problem
\begin{equation}
    \min_{\mathbf{I}} f_{QS} + w_{R} f_{R} + w_{C} f_{C} + w_{\iota} f_{\iota},
    \label{eq:epsilon-constraint-optimization}
\end{equation}
where $w_{R}$, $w_{C}$, and $w_{\iota}$ are chosen to be sufficiently large to dominate the objective function unless they reach their target value within $\approx0.1 \%$. Other studies \cite{gil2026,gil2026a} have developed improvements to this approach using an augmented Lagrangian formulation to more robustly enforce constraints and could also be applied here. Our approach to identify optimal solutions is then to solve \cref{eq:epsilon-constraint-optimization} for a range of values of $\iota_T$ and $I^{p=20}_T$ at fixed $V_T$. We initialize the optimization from the selected sampled geometries identified in \cref{fig:stage-II-pareto-front}. We did not find a significant difference between the $N_{\theta} = 10$ and $11$ initial conditions, so we focus on the $N_{\theta} = 10$ results here.

We employ an adaptive resolution approach. At fixed $I^{p=20}_T$, $V_T$, and $\iota_T$, we solve \cref{eq:epsilon-constraint-optimization} starting from the two-stage initial condition with $\mathit{ntor} = \mathit{mpol} = 6$ and a modest residual target. We then increase the maximum mode numbers to $9$ and then $12$, initializing from the previous converged solution and decreasing $R_T$ each time down to $R_T \approx 5\times10^{-6}$, which we empirically determined to be sufficient for nested flux surfaces \cite{giulianiDirectStellaratorCoil2023} using Poincar\'e plots. This continuation addresses the numerical challenges known to occur when starting the optimization from a high-resolution equilibrium \cite{jang2026}.

We repeat the $\iota$ scan at $I^{p=20}_T \in {100, 150, 200, 250}~\unit{\kilo\ampere}$. We choose these values using the force model described in \cref{sec:design-specification}: $\qty{250}{\kilo\ampere}$ represents our estimated upper bound on feasible currents, and we will show that even this value remains sufficiently below the HTS force tolerance---indicating that higher currents may be accessible and that $\qty{100}{\kilo\ampere}$ is a highly conservative lower bound which still allows access to reasonable physics targets. This scan accounts for uncertainty in the approximate force model while showing the device's flexibility across operating conditions.

\subsubsection{QS Error, Transform, and Current Limits}

\begin{figure}[H]
    \centering
           \includegraphics[width=\textwidth]{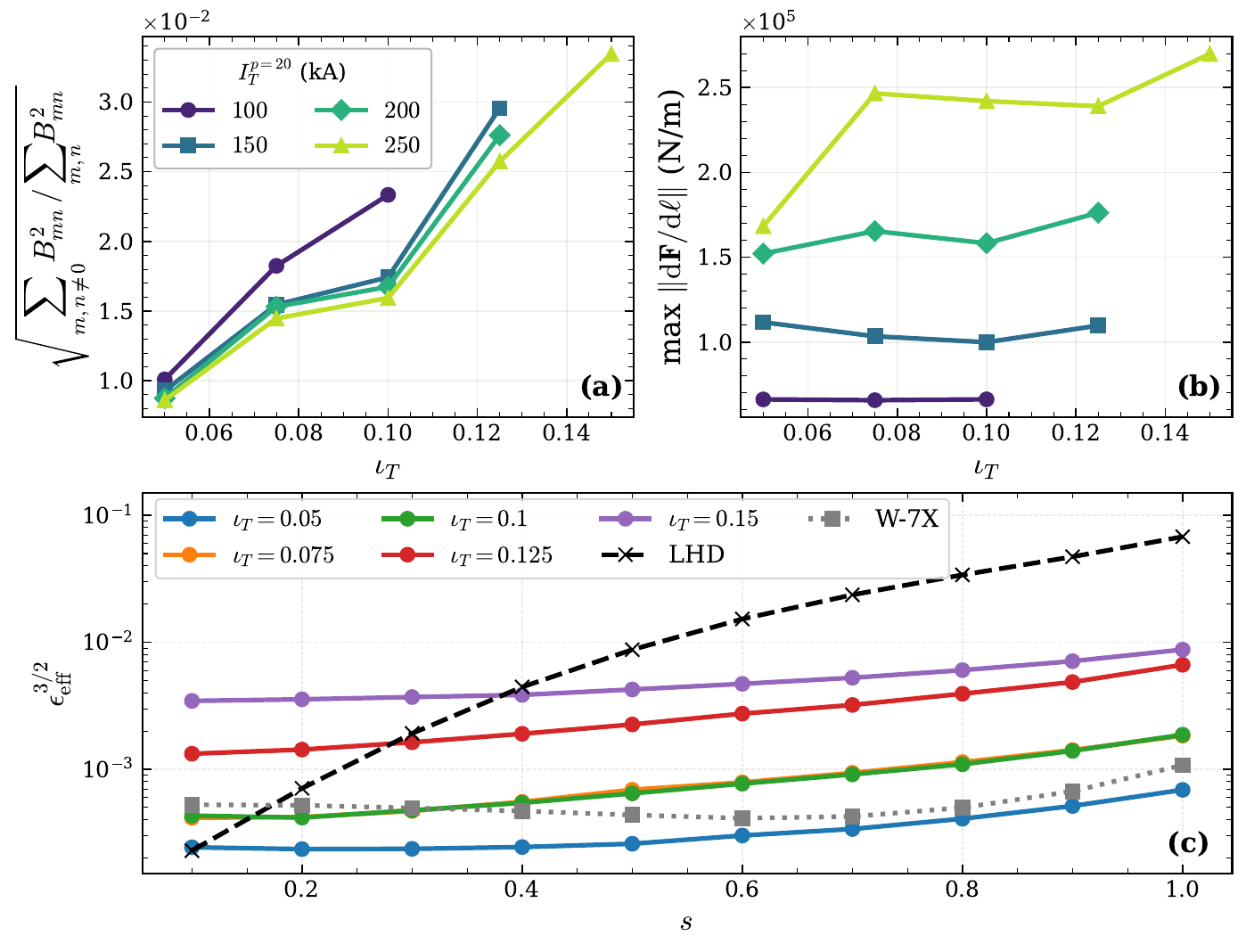}  
    \caption{(a) Square root of the normalized sum of quasi-axisymmetry-breaking Fourier harmonics in Boozer coordinates for single-stage optimizations at varying rotational transform $\iota$ and p-norm targets $I^{p=20}_T$, starting from the $N_{\theta} = 10$ initial condition in \cref{fig:stage-II-pareto-front}. Each p-norm threshold is shown as a different marker shape. We only plot optimizations where full nested flux surfaces were observed. (b) The same scan, instead plotting the maximum force per unit length in the coil set. (c) Effective ripple $\epsilon_{eff}^{3/2}$ versus normalized toroidal flux $s$ for the $\iota_T = 0.05$--$0.15$ configurations at $I^{p=20}_T = \qty{250}{\kilo\ampere}$ from panel (a) compared to both an inward-shifted LHD \cite{murakami2002} and $\beta=2\%$ standard W-7X configuration.}
    \label{fig:single-stage-scan}
\end{figure}

We plot the results of our vacuum single-stage optimization scan with fixed $V_T = \qty{0.3}{\cubic\meter}$ in \cref{fig:single-stage-scan}. In \cref{fig:single-stage-scan}(a), we again plot the QS error expressed as the square root of the normalized sum of quasi-axisymmetry-breaking modes in Boozer coordinates on the surface. In \cref{fig:single-stage-scan}(b), we plot the maximum pointwise force $|d\mathbf{F}/dl|$ in each coil set of the scan, computed using the analytic self-force model described in \cref{sec:design-specification}. In both panels, we show different $I^{p=20}_T$ as various markers (circles, squares, diamonds, and triangles for increasing current threshold). There is a clear tradeoff between rotational transform, QS error, and peak coil currents, as expected from our two-stage analysis. At the smallest $\iota_{T} = 0.05$, the QS errors are all approximately the same, indicating a lower bound on the QS error achievable by this coil configuration at this rotational transform and residual target, and that larger currents do not provide additional benefit. As $\iota_T$ is increased at a constant current threshold, the QS error increases for all values of $I^{p=20}_T$, but worsens more rapidly for smaller current thresholds, indicating the benefit of larger currents at higher $\iota$. 

In \cref{fig:single-stage-scan}(b), we find that all design points remain below the HTS force tolerance of $4$–$8\times10^5~\unit{\newton\per\meter}$ established in \cref{sec:design-specification}; the maximum is $\approx2.75 \times 10^5~\unit{\newton\per\meter}$, at most $70\%$ of the minimum limit in that range. We reemphasize that these estimates assume idealized circular cross-sections and an analytic self-force model, and the absolute values should be treated as order-of-magnitude indicators rather than reliable engineering limits. At fixed $I^{p=20}_T$, the maximum pointwise force $|d\mathbf{F}/dl|$ varied little across optimizations, despite differing current distributions between them. This confirms that the p-norm captures the dominant control over coil forces and that the precise distribution of current across coils has only minor effects at fixed current thresholds. The lower force for the $\iota=0.05$, $I^{p=20}_T = \qty{250}{\kilo\ampere}$ case arises because the optimizer converged to $I^{p=20} \approx \qty{202}{\kilo\ampere}$---i.e., the current constraint was not binding, consistent with the QS floor already being reached at this low $\iota$.  Unlike reactor-scale designs where large toroidal fields push dipole coils near their HTS limits \cite{kaptanogluReactorscaleStellaratorsForce2025}, the binding constraints here are more likely to arise from torques on the support structure and the number of HTS turns required to meet the current targets in the smaller coil geometry, both of which motivate the detailed structural analysis deferred to future work.

In \cref{fig:single-stage-scan}(c), we directly evaluate the confinement quality of the device by calculating the effective ripple $\epsilon_{eff}^{3/2}$ \cite{unalmis2026} for the $I^{p=20}_T = \qty{250}{\kilo\ampere}$ scan and compare it to both a neoclassically optimized, inward-shifted LHD \cite{murakami2002} and a $\beta = 2\%$ standard W-7X configuration as a function of the toroidal flux $s$. As expected, the effective ripple correlates strongly with the QS error metric of \cref{fig:single-stage-scan}(a), increasing with larger $\iota_T$. Our optimized configurations with $\iota_T < 0.1$ all have comparable transport properties to the W-7X equilibrium, and the highest $\iota_T$ cases still demonstrate better average confinement than the LHD configuration over the volume. While not achieving the level of precise confinement found in some magnetic configurations \cite{landremanMagneticFieldsPrecise2022}, this analysis demonstrates that these configurations have sufficient confinement relative to existing experiments.

In \cref{fig:single-stage-scan}, we only plot the subset of scanned configurations that met the $\iota$, current, and residual constraints, and subsequently had nested flux surfaces. This subset defines the accessible parameter range explored in this study. As with QS error, we find that increasing the current threshold allows us to achieve higher $\iota$, with $\qty{100}{\kilo\ampere}$ reaching $0.1$ while $\qty{250}{\kilo\ampere}$ can reach $0.15$. Notably, the vacuum $\iota$ achievable by each of these current thresholds is physically relevant for stabilizing MHD instabilities in hybrid scenarios, as we will discuss in \cref{subsec:finite-beta-optimization}. 

\begin{figure}[H]
    \centering
           \includegraphics[width=\textwidth]{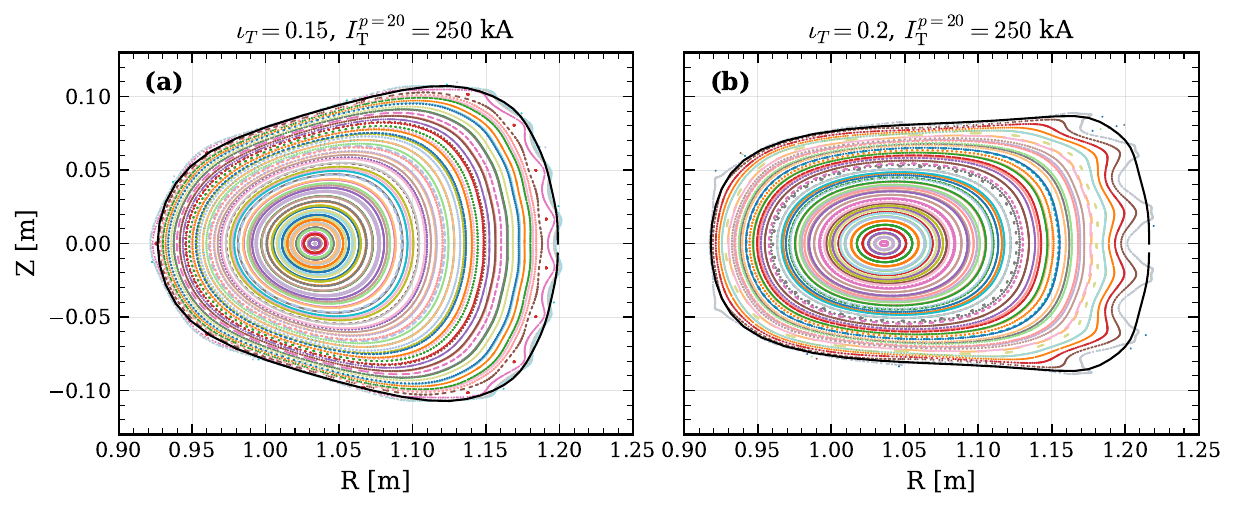}
    \caption{Poincar\'e plots at $\phi = 0$ for the (a) $\iota_T = 0.15$ and (b) $\iota_T = 0.2$ configurations, both with $I^{p=20}_T = \qty{250}{\kilo\ampere}$ and $V_T = \qty{0.3}{\cubic\meter}$. The optimized surface boundary is shown in black.}
    \label{fig:poincare-plots}
\end{figure}

\Cref{fig:poincare-plots} contrasts a successful case at $\iota_T = 0.15$ with a rejected case at $\iota_T = 0.2$ where the residual threshold was broken, both at $I^{p=20}_T = \qty{250}{\kilo\ampere}$ and $V_T = \qty{0.3}{\cubic\meter}$. While the former shows nested flux surfaces throughout the entire volume aside from minor edge rippling, the latter exhibits large non-nested regions on the outboard side near the coils, validating our choice of $R_T$. Throughout this scan, we observed that islands or chaotic regions typically formed on the outboard side, likely a result of its closer distance to the coil set and strongly varying local fields there.

We caution that a rejected scan point does not necessarily mean the combination of $\iota_T$, $I^{p=20}_T$, and $V_T$ is unachievable for this coil set. In some optimizations, we observed the onset of chaotic regions even as $R$ decreased, indicating some equilibria are more sensitive to field errors than others in a way not factored into our optimization routine. The accessible operational space could be expanded with an objective that directly targets robustness of the flux surfaces to residual field error and we leave this to future work.

\subsubsection{Geometric Interpretation of the Tradeoffs}

\begin{figure}[H]
    \centering
           \includegraphics[width=\textwidth]{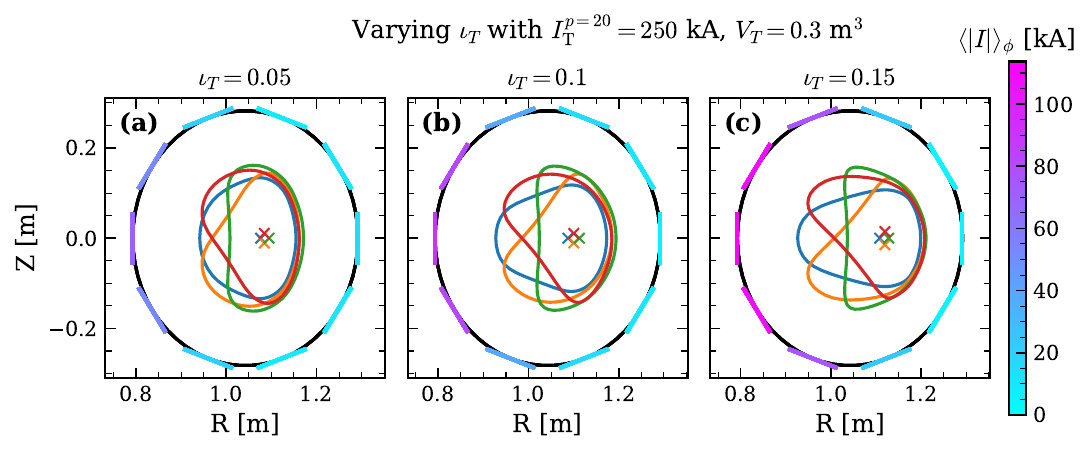}\\
           \includegraphics[width=\textwidth]{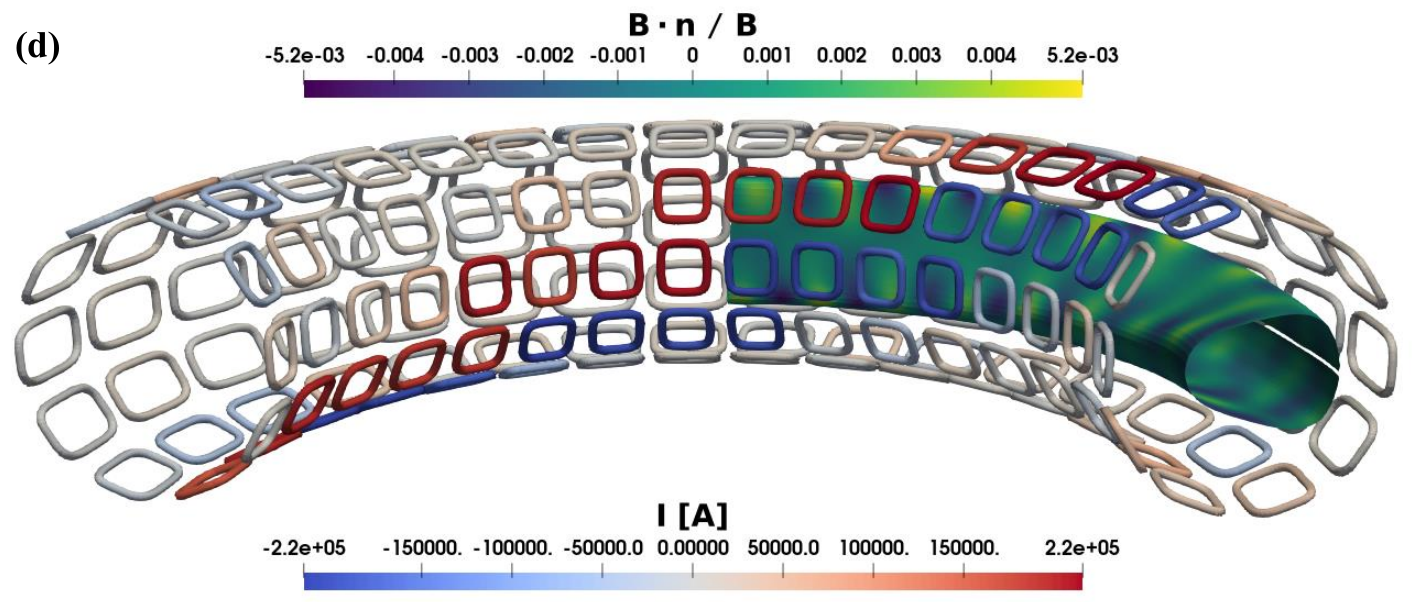}
    \caption{Optimized magnetic surface cross sections at equally spaced toroidal locations in a field period, with the magnetic axis location for each shown as an X in the corresponding color. We scan $\iota_T = 0.05, 0.1, 0.15$ from (a)-(c), respectively, with fixed $I^{p=20}_T = \qty{250}{\kilo\ampere}$ and $V_T = \qty{0.3}{\cubic\meter}$. The vacuum vessel is shown in black, and the dipole coil cross sections are colored based on their toroidally averaged current magnitude. (d) 3D representation of the configuration in (c), with a half field period of the plasma surface colored by the normalized normal field $\mathbf{B}\cdot\mathbf{n}/B$ and a field period of the dipole coils colored by their currents. \label{fig:iota-cross-sections}}
   \end{figure}

 For a deeper look into the tradeoffs in \cref{fig:single-stage-scan}, we now plot the magnetic surface cross sections, axis locations, and toroidally averaged currents for some representative cases. In \cref{fig:iota-cross-sections}, we fix $I^{p=20}_T = \qty{250}{\kilo\ampere}$ with increasing $\iota_T$ from $0.05$ to $0.15$, and include a 3D rendering of the $\iota_T = 0.15$ solution in \cref{fig:iota-cross-sections}(d). As the rotational transform increases from \cref{fig:iota-cross-sections}(a) to \cref{fig:iota-cross-sections}(c), the cross section becomes more highly shaped, while the level of axis torsion remains roughly constant. This matches both the near-axis expectations of \cref{fig:inverse-envelope-volume-ratio} and the stage-I optimization of \cref{fig:stage-I-eq}.

The added shaping requires a slight increase of volume for the axisymmetric envelope enclosing the plasma. Computing the minimum axisymmetric envelope enclosing the surface over a field period, we find that the volume ratio of the plasma to the envelope decreases from $V/V_{env} = 0.75$ to $0.68$ to $0.64$ across (a)--(c), in qualitative agreement with the near-axis toy model of \cref{sec:near-axis-tradeoffs}. At fixed $V_T$ the envelope must therefore grow, but it does so almost entirely on the outboard side where the coil currents are the smallest to limit the increase in residual until it loses nested flux surfaces, consistent with \cref{fig:poincare-plots}. This follows directly from the field-error/current tradeoff of \cref{subsec:stage-II}: the maximum coil current limits how far the surface can sit from the coils, while the per-coil poloidal ripple sets a minimum field error that grows as the surface approaches them. Together these effects bound the coil-surface distance from both sides, limiting the allowable volume of the axisymmetric envelope and controlling the type of shaping that can be achieved within it at the cost of QS error. This also explains why the optimizer failed to satisfy all constraints for larger $\iota_T$/smaller $I^{p=20}_T$, as no surface within the envelope could meet the residual, current, and transform targets at any level of QS error.

Furthermore, the inboard coils carry the largest currents and provide the majority of the shaping in all cases, consistent with the inboard-localized shaping preferred by these QA equilibria. As $\iota$ increases, the toroidally averaged current increases for both the inboard poloidal coil array and its neighbors. This is a byproduct of enforcing a limit on the maximum current, as all of these cases have approximately the same maximum current, but as we increase $\iota$, the proportion of the coils which are at or near this maximum increases in order to provide the necessary transform. Importantly, however, we find that the outboard coils have small currents in all of the configurations of this scan, which we leverage to promote sparsity in \cref{subsec:sparsity}. The corresponding 3D view in \cref{fig:iota-cross-sections}(d) further visualizes this effect, as the dipole currents concentrate into an inboard banana-like pattern \cite{hennebergCompactStellaratortokamakHybrid2024a}. The residual normal field $\mathbf{B}\cdot\mathbf{n}/B$ on the boundary remains small, with structure set by the discrete coil spacing rather than large-scale field errors.

\begin{figure}[H]
\centering
       \includegraphics[width=\textwidth]{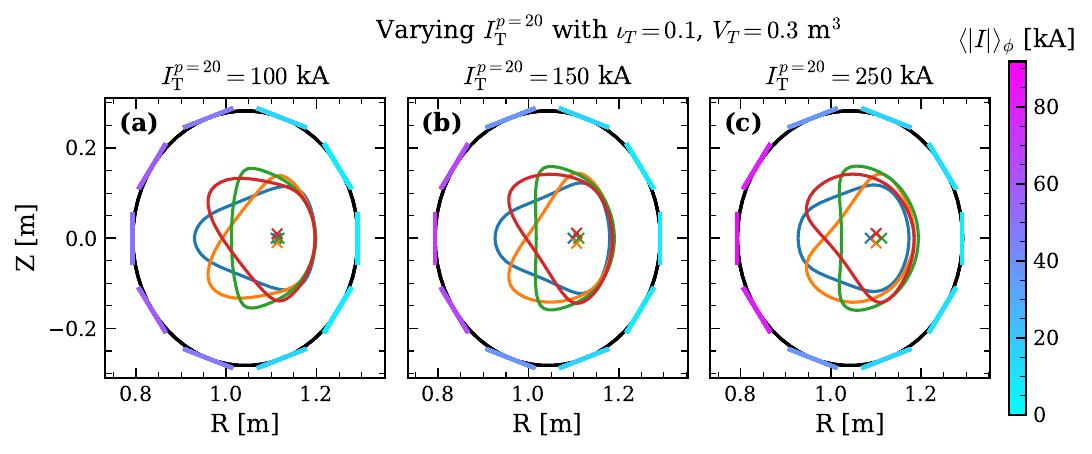}
\caption{Optimized magnetic surface cross sections at equally spaced toroidal locations in a field period, with the magnetic axis location for each shown as an X in the corresponding color. We scan $I^{p=20}_T = 100, 150, \qty{250}{\kilo\ampere}$ from (a)-(c), respectively, with fixed $\iota_T = 0.1$ and $V_T = \qty{0.3}{\cubic\meter}$. The vacuum vessel is shown in black, and the dipole coil cross sections are colored based on their toroidally averaged current magnitude.\label{fig:current-cross-sections}}
\end{figure}

\Cref{fig:current-cross-sections} shows the same quantities but for configurations with increasing $I^{p=20}_T$ and fixed $\iota_T = 0.1$. In this scan, the axisymmetric envelope is nearly constant, with the volume ratio $V/V_{env} = 0.68 \pm 0.004$ across (a)--(c). This finding supports the conclusion that the current threshold acts primarily on how the required shaping is distributed within the envelope and not the size itself, degrading QS error as the current limit tightens. For the minimum current configuration, \cref{fig:current-cross-sections}(a), the optimizer reduces axis torsion to near-zero, observed via the roughly constant magnetic axis locations. The optimizer preferentially reduces torsion because a more planar axis minimizes the maximum distance from the coils to the plasma boundary, requiring less current to produce the same shaping field. However, near the magnetic axis it can be shown that torsion is required for quasi-symmetry with finite rotational transform \cite{landremanDirectConstructionOptimized2018}. As this source of QS-compatible rotational transform is removed, the remaining $\iota_T$ must be supplied by rotating ellipticity instead, making the equilibria more elongated, with non-axisymmetric perturbations spread over a wider poloidal extent and driving up the QS error. We can also see this effect via the ratio of the inboard to outboard toroidal currents, which is much smaller at lower current limits. As the current limit becomes even smaller than that shown here, we find rotating-ellipse-like configurations which represent the minimum-current path to a fixed rotational transform, but at a cost to QS quality.

While the primary intent of running with multiple values of $I^{p=20}_T$ was to demonstrate the capability to reach desired rotational transform at a range of possible engineering constraints, it can also hint at the flexibility of the design. For example, in \cref{fig:current-cross-sections}(a)-(c) we demonstrated the delocalization of the perturbation when more strongly penalizing the current, analogous to the contrast between the inboard-localized banana pattern of \cref{fig:iota-cross-sections}(d) and the helical coils of torsatrons \cite{hartwellDesignConstructionOperation2017a}. The axisymmetric dipole coil array therefore offers broad flexibility, accessing a wide range of vacuum $\iota$ while maintaining low QS error and spanning a variety of equilibrium shapes.

\begin{figure}[H]
    \centering
           \includegraphics[width=\textwidth]{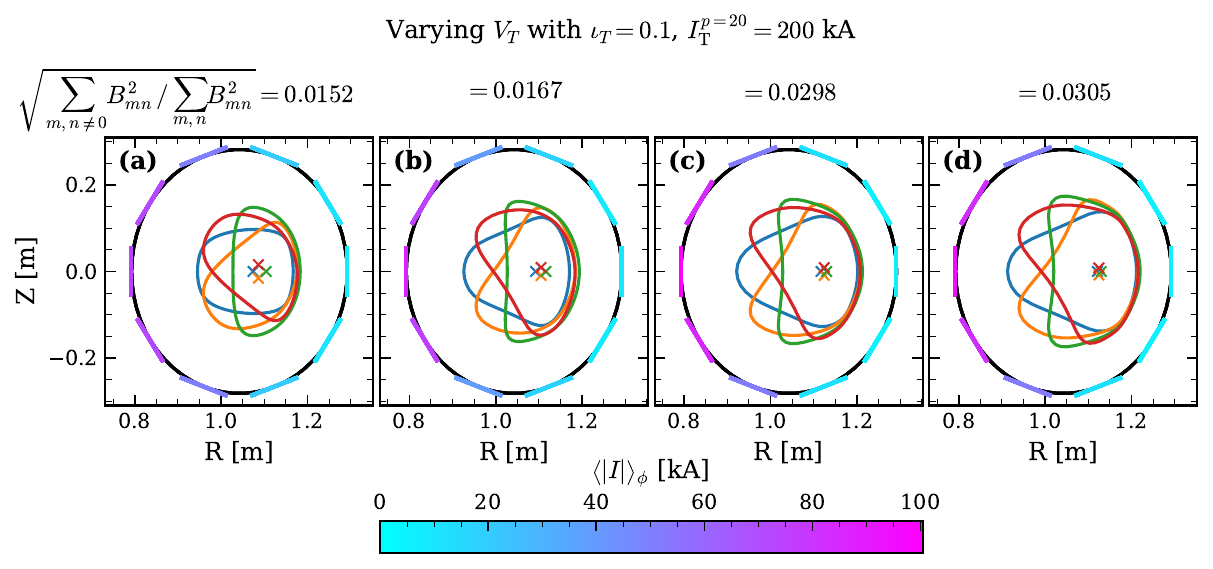}
    \caption{Configurations with a varying volume target of (a) $\qty{0.25}{\cubic\meter}$ (b) $\qty{0.3}{\cubic\meter}$ (c) $\qty{0.35}{\cubic\meter}$ and (d) $\qty{0.4}{\cubic\meter}$ at fixed $I^{p=20}_T = \qty{200}{\kilo\ampere}$ and $\iota_T = 0.1$. We show the cross sections at $4$ equally spaced toroidal locations for a field period, along with their magnetic axis location as an X in the corresponding color. The vacuum vessel is shown in black, and the dipole coil cross sections are colored based on their toroidally averaged current magnitude. We plot the same QS error metric for each configuration as in previous figures above the subplots. \label{fig:single-stage-volume-scan}}
   \end{figure}

All single-stage optimizations in \cref{fig:single-stage-scan} were at a fixed volume target $V_T = \qty{0.3}{\cubic\meter}$. To explore the effects of varying this target, we plot a scan of increasing $V_T$ in \cref{fig:single-stage-volume-scan}. We fix $\iota_T = 0.1$ and $I^{p=20}_T = \qty{200}{\kilo\ampere}$, and again plot the cross sections, axis location, and toroidally averaged currents. We again stop the scan when we cannot find a configuration with nested flux surfaces, which occurred at $\qty{0.45}{\cubic\meter}$ for this case.

In contrast to \cref{fig:current-cross-sections}, the axisymmetric envelope is not invariant across this scan. Because $\iota_T$ is fixed, the volume ratio in the envelope is nearly constant with $V/V_{env} = [0.67, 0.66, 0.69, 0.71]$ across (a)--(d), so the plasma volume can only increase if the axisymmetric envelope does as well. The modest gain in volume ratio is paid for by an increase in QS error, which doubles between the $V = \qty{0.25}{\cubic\meter}$ and $V = \qty{0.4}{\cubic\meter}$ cases. The envelope grows by moving the boundary toward the coils, predominantly on the outboard side where the coil currents are the smallest. Above $\qty{0.4}{\cubic\meter}$, the envelope required at this $\iota_T$ pushes the boundary too close to the coils, breaking the residual threshold and losing nested flux surfaces. Similar to the low current cases, we find that the maximum volume solution coincides with an equilibrium with little to no axis torsion.

Together, the analyses in this section demonstrate the physical mechanisms creating the competition between volume, $\iota$, current, and QS error. In order to have a small residual and nested flux surfaces, there is a minimum allowable coil-plasma distance which limits the available volume of the axisymmetric envelope confining the equilibrium. Within that limit, the fraction of the envelope that a quasi-axisymmetric surface can fill is set primarily by $\iota$ (\cref{sec:near-axis-tradeoffs}). This creates direct competition between the $\iota$ and volume targets, while the current threshold governs how efficiently that shaping can be delivered, with QS error serving as the common tradeoff variable. Eventually, even degrading QS quality is insufficient, and no configuration can meet the combination of targets. As a direct example of this interplay, we found that repeating our scan of \cref{fig:single-stage-scan} at only a modest decrease in volume to $V_T = \qty{0.25}{\cubic\meter}$ increased the achievable $\iota$ to $0.15$ for $I^{p=20}_T = \qty{200}{\kilo\ampere}$ and $0.2$ for $I^{p=20}_T = \qty{250}{\kilo\ampere}$, up from $0.125$ and $0.15$ at $V_T = \qty{0.3}{\cubic\meter}$, respectively.

\subsubsection{Sparse Coil Arrays}
\label{subsec:sparsity}

As a final improvement to our vacuum single-stage optimization routine, we perform a preliminary investigation into incorporating sparsity into the dipole coil array. A dense array of HTS coils will severely limit access to the device, as it requires that the entire surface of the vacuum vessel be kept at cryogenic temperatures. This makes diagnostics infeasible, defeating the advantages of the experiment. Clearly, some level of sparsity in the coil array needs to be incorporated to allow accessibility.

Several previous studies have investigated sparsity in related designs using either dipole coils \cite{wuPlanarCoilOptimization2025} or permanent magnets \cite{kaptanogluPermanentMagnetOptimizationStellarators2022, kaptanogluGreedyPermanentMagnet2023}. The premise of these routines is removing individual coils which do not contribute significantly to reducing the magnetic field error on the surface, typically through the inclusion of the L1 or L0 norm to penalize small or non-zero coil currents, respectively. However, this approach is less effective for this design, as it breaks the axisymmetry of the coil set and the benefits that come with it. Instead, we incorporate sparsity in terms of its poloidal extent, assuming the gaps in the array also remain axisymmetric. As an example, we take the same $N_{\theta} = 10$ configuration and remove the outboard midplane coil, opening a gap of approximately $\qty{20}{\centi\meter}$ between coil filaments in the poloidal direction at all toroidal locations (in between TF coils). This choice was motivated by both analytic predictions/coil generation of quasi-axisymmetric equilibria with only inboard shaping \cite{plunkQuasiaxisymmetricMagneticFields2018, plunkPerturbingAxisymmetricMagnetic2020,hennebergCompactStellaratortokamakHybrid2024a,hennebergVarietyCoilSets2025} and the observation that all of our previous optimized configurations have the largest coil currents on the inboard side. This configuration is the same as that visualized in \cref{fig:coil_geometry} with the green coils removed.

\begin{figure}[H]
    \centering
           \includegraphics[width=\textwidth]{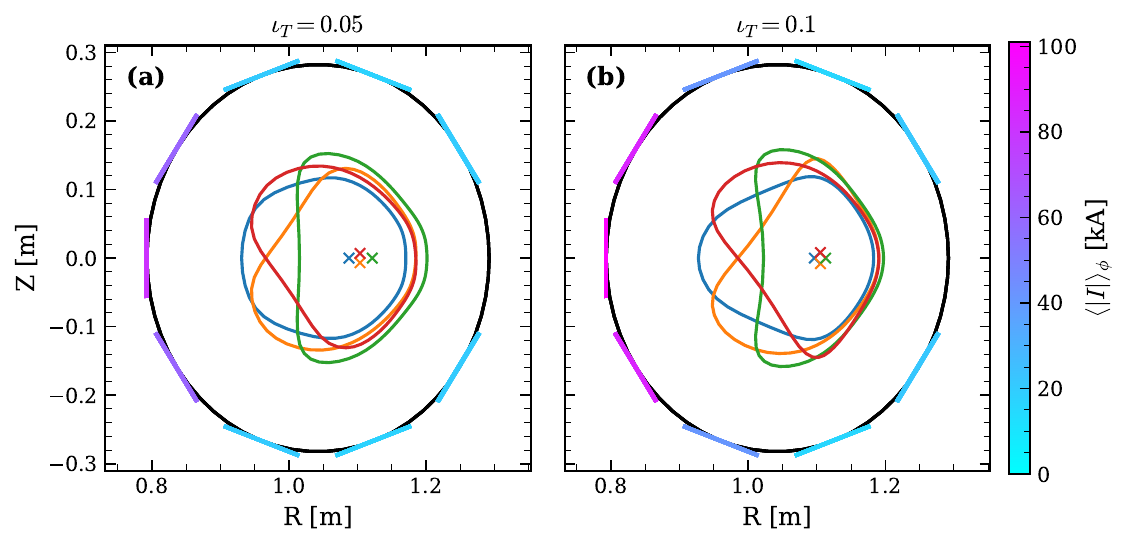}
    \caption{Magnetic surface cross sections at equally spaced toroidal locations for a field period for the outboard sparsity configuration with $I^{p=20}_T = \qty{250}{\kilo\ampere}$, $V_T = \qty{0.3}{\cubic\meter}$, and (a) $\iota_T = 0.05$ and (b) $\iota_T = 0.1$. The magnetic axis locations are shown as X's colored to match their cross section, the vacuum vessel is shown in black, and the dipole coil cross sections are colored based on the toroidally averaged current magnitude. \label{fig:single-stage-outboard-sparsity}}
\end{figure}

We then optimize the coil currents using the same single-stage optimization routine as before, initializing from the dense array optimized solutions of \cref{fig:single-stage-scan}. We plot the result in \cref{fig:single-stage-outboard-sparsity}, showing only the scan for $I^{p=20}_T = \qty{250}{\kilo\ampere}$, as we have demonstrated that this threshold is likely to be within engineering limits. We show $\iota_T = 0.05$ and $0.1$ to demonstrate that we can still achieve sufficient transforms in the sparse configuration. We find that the sparse configurations converge to surfaces similar to those found with the dense array in \cref{fig:iota-cross-sections}. In both configurations, we find that $f_{QS}$ becomes only $10-20\%$ worse than the dense configuration, indicating that this sparse array retains QS quality comparable to the dense array we assessed via $\epsilon_{eff}$ in \cref{fig:single-stage-scan}(c). Ultimately, the actual sparse configuration will require further iteration to optimize the tradeoff between adding further sparsity to promote accessibility and remaining flexible in achieving many magnetic configurations. However, it is clear that the single-stage optimization routine provides a robust method for this investigation, as it facilitates obtaining equilibria consistent with any specified coil layout.

\subsection{Finite-$\beta$ Optimization}
\label{subsec:finite-beta-optimization}

We next include finite toroidal plasma current in the optimization routine. The single-stage optimization approach employed in the previous section only applies to vacuum stellarator equilibria. To assess the experiment as a hybrid tokamak-stellarator, we must consider non-axisymmetric equilibria with inductively driven toroidal current. When considering equilibria with non-zero pressure and current profiles, or finite-$\beta$ equilibria, we must solve for the full ideal MHD equilibrium using existing codes, one example being VMEC \cite{hirshmanSteepestdescentMomentMethod1983}. Prior work has integrated VMEC into single-stage optimization routines \cite{jorgeSinglestageStellaratorOptimization2023, baillodIntegratingNovelStellarator2025}, where the equilibrium and coil terms are coupled via the quadratic flux on the surface, and derivatives are obtained analytically from the coils and using finite differences for the equilibrium. However, it has been found to be less robust than the vacuum approach \cite{baillodIntegratingNovelStellarator2025}, and we further show here that we do not find it necessary to obtain sufficiently optimized finite-$\beta$ equilibria for our design.

\begin{figure}[H]
\centering
        \includegraphics[width=\textwidth]{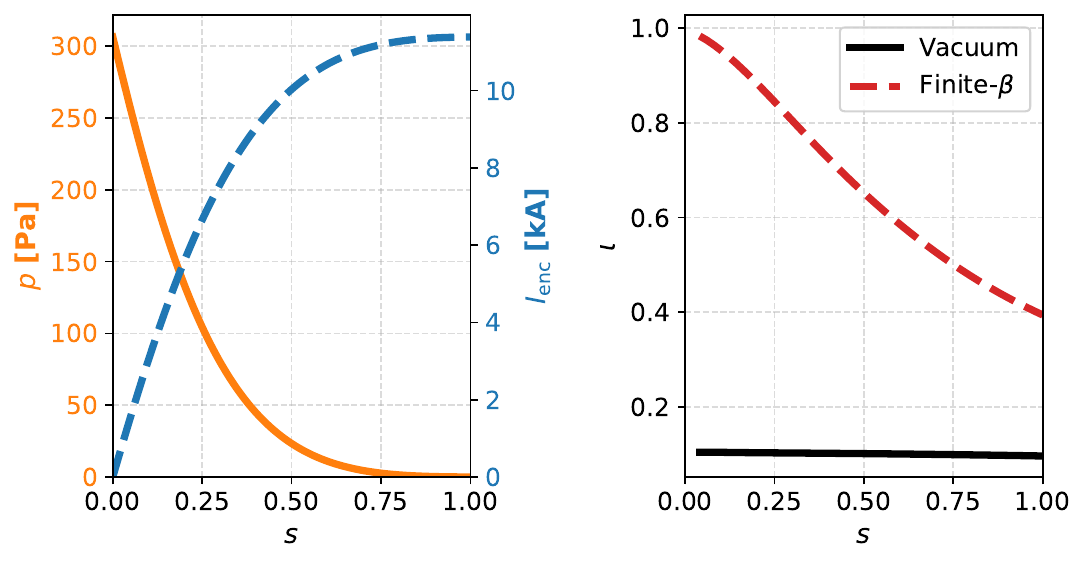}
\caption{Left: realistic HBT-EP profiles we incorporated into our vacuum solution from the single-stage optimization routine as a function of the normalized toroidal flux $s$. The pressure profile $p$ is shown in solid orange and the total enclosed toroidal current $I_{enc}$ is shown in dashed blue. Right: rotational transform profile $\iota$ for both the vacuum equilibrium in solid black and finite-$\beta$ VMEC equilibria in dashed red. \label{fig:finite-beta-profiles}}
\end{figure}

For this analysis, we add realistic plasma profiles from the HBT-EP device with a pressure of $p \approx \qty{306}{\pascal}$ on axis and an ohmically-driven toroidal current $I_{enc} = \qty{11.3}{\kilo\ampere}$. These profiles correspond to an on-axis $\beta = 0.33\%$ and volume-averaged $\langle \beta \rangle = 0.07\%$, and we therefore assume zero bootstrap current. We solve for the ideal MHD equilibrium using VMEC in fixed-boundary mode with these profiles and the surface boundary from the $\iota = 0.1, I^{p=20}_T = \qty{150}{\kilo\ampere}$ case in \cref{fig:single-stage-scan}. In \cref{fig:finite-beta-profiles}, we plot the HBT-EP input profiles and the resulting rotational transform $\iota$ both for the vacuum and finite-$\beta$ equilibria.

This example is highly relevant for hybrid operation with an on-axis $\iota \approx 1$, matching that of tokamaks, and an edge external rotational transform fraction of $f_{ERT} = \iota_{vac}(a) / \iota(a) \approx 0.25$. This level of vacuum transform has been both theoretically predicted \cite{fuVerticalStabilityCurrentcarrying2000, kuNonaxisymmetricShapingTokamaks2009} and experimentally observed \cite{pandyaLowEdgeSafety2015,li2026} to stabilize MHD instabilities in current-carrying stellarators. Specifically, \cite{pandyaLowEdgeSafety2015} found that an edge $f_{ERT} = 0.1$ achieved disruption-free operation, making this design point well past the observed level to achieve these goals. While detailed MHD stability modeling is beyond the scope of this initial feasibility study and is prioritized for future work, we have verified this equilibrium is Mercier stable using VMEC, which, when combined with the prior experimental evidence, makes this equilibrium a promising candidate for hybrid operation.

\begin{figure}[H]
\centering
        \includegraphics[width=\textwidth]{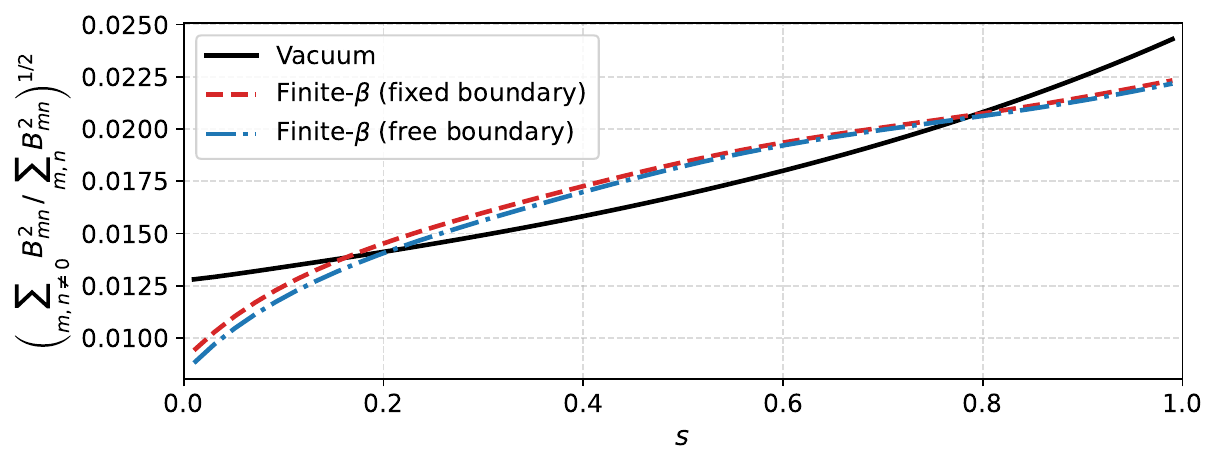}
\caption{QS error expressed as the square root of the sum of quasi-axisymmetry-breaking to total Fourier harmonics ratio in Boozer coordinates as a function of the normalized toroidal flux $s$, for the vacuum equilibrium in black, fixed boundary finite-$\beta$ equilibrium in red, and free boundary finite-$\beta$ equilibrium in blue. The equilibria are the same as those from \cref{fig:finite-beta-profiles}. \label{fig:finite-beta-qs-error}}
\end{figure}

Adding these profiles changes the equilibrium field, so the vacuum QS error and dipole-coil currents must be reassessed. To address the first concern, we take the VMEC equilibria from \cref{fig:finite-beta-profiles}, both with and without the plasma profiles, transform them into Boozer coordinates, and calculate the quasi-axisymmetry-preserving/breaking Fourier harmonics of the magnetic field. We plot the result of this calculation in \cref{fig:finite-beta-qs-error}. Although the finite-$\beta$ equilibrium is not re-optimized for QS, the profiles only weakly perturb the QS error profile, which is even slightly improved at some radial locations. This finding is most likely a result of both the relatively small $\beta$ from these profiles and targeting only modest QS in our equilibria for increased flexibility in the experiment, and it is probable that this outcome does not hold for equilibria with high $\beta$ or QA at the precise levels obtained in some stage-I equilibria \cite{landremanMagneticFieldsPrecise2022}.

From an optimization standpoint, the QS error staying approximately the same implies that we do not need to reoptimize the equilibrium through stage-I or the finite-$\beta$ single-stage approach. Instead, we can just optimize the coil currents through another stage-II optimization, this time considering the magnetic field contribution from the plasma currents as well. This problem is formulated in a similar fashion to \cref{eq:stage-II-objective-function},
\begin{equation}
    \min_{\mathbf{I}} \int_S \left| \left(\mathbf{B}(\mathbf{I}) + \mathbf{B}_{TF} + \mathbf{B}_{p}\right)\cdot \hat{\mathbf{n}} \right|^2 dS,
    \label{eq:finite-beta-objective-function}
\end{equation}
where we include the normal magnetic field on the equilibrium surface due to the plasma currents, $\mathbf{B}_{p} \cdot \hat{\mathbf{n}}$, using the virtual casing principle \cite{lazersonVirtualcasingPrinciple3D2012} and we again find regularization unnecessary for a well-conditioned solution. After this optimization, the maximum individual coil current decreases by around $20\%$ compared to the vacuum solution due to the inclusion of the magnetic field from the plasma current. Finally, we run VMEC in free-boundary mode to ensure an equilibrium consistent with the new optimized coil currents. We plot the free boundary QS error in \cref{fig:finite-beta-qs-error}, showing that it also remains approximately the same as the vacuum case.

While we conclude the optimization here, this configuration could be further improved in the future by including poloidal field coils \cite{hennebergCompactStellaratortokamakHybrid2024a} or passing this configuration into a finite-$\beta$ single-stage algorithm, which likely will only require a small number of iterations to converge due to its already near-optimal levels.

\section{Benefits of Dipole Coils on Tokamak Operation}
\label{sec:tokamak-benefits}

Up to this point, we have only considered the benefits of dipole coils for creating non-axisymmetric equilibria. However, one of the main advantages of the axisymmetric dipole coil array is that it can also be exploited during tokamak operation. In this section, we discuss two primary benefits of dipole coils on tokamak operation which occur when aligning the coil currents in either the poloidal or toroidal direction---correcting TF coil ripple and creating shaped tokamak equilibria, respectively.

\subsection{Correcting TF Ripple}
\label{subsec:tf-ripple}
Tokamaks are nominally axisymmetric, but a finite number of TF coils introduces magnetic field ripple. If sufficiently large, this ripple degrades confinement by trapping particles in local magnetic wells or modifying banana-trapped particle orbits \cite{wessonTokamaks2011}. These effects typically set a lower bound on the number of TF coils that can be used, quantified by the parameter \cite{kripnerToroidalMagneticField2023}
\begin{equation}
    \delta = \frac{B_{max} - B_{min}}{B_{max} + B_{min}}\bigg|_\phi,
    \label{eq:tf-ripple-parameter}
\end{equation}
where $\delta$ is typically constrained to be on the order of $1 \%$ at the outboard midplane and $0.01 \%$ at the axis \cite{wessonTokamaks2011}, though these are only approximations and are set by the specific transport properties of each device. The ripple reaches a maximum at the outboard midplane, and this is where its effects are typically quantified \cite{devriesEffectToroidalField2008, scott2020, kripnerToroidalMagneticField2023}.

\begin{figure}[H]
 \centering
        \includegraphics[width=\textwidth]{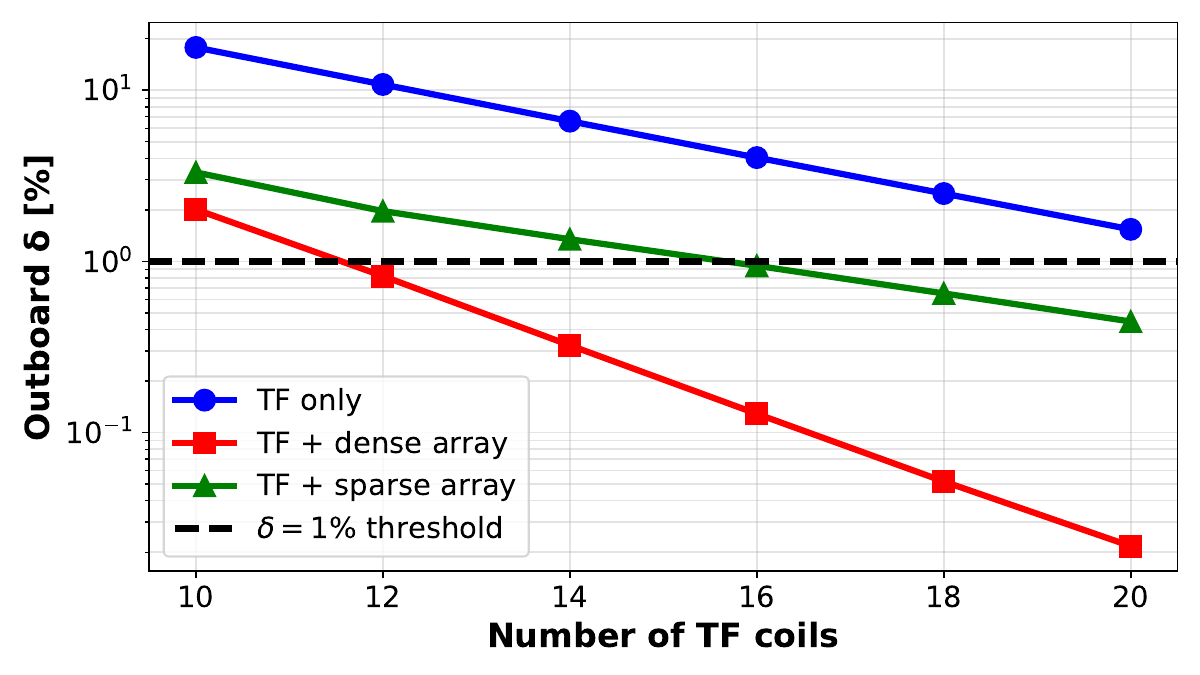}
 \caption{Coil ripple parameter $\delta$ (\cref{eq:tf-ripple-parameter}) on the outboard midplane of a sample tokamak configuration as a function of the number of TF coils for several cases: the ripple from the TF coils alone (blue circles), the ripple from the TF coils and a dense dipole coil array (red squares), and the ripple from the TF coils and dipole coils without the outboard midplane coil (green triangles). We fix the number of toroidal dipole coils to $N_{\phi} = 2 \times N_{TF}$. \label{fig:tf-ripple-correction}}
\end{figure}

The axisymmetric dipole array can correct this ripple, allowing for the use of fewer TF coils and improving access on the outboard side of the plasma. Variations of this idea have been proposed for use in both tokamaks and stellarators \cite{boozerNeededComputationsComputational2024, elderCurrentPotentialPatches2024, baillodEnhancingStellaratorAccessibility2025a}. \Cref{fig:tf-ripple-correction} shows the TF coil ripple parameter $\delta$ versus the number of TF coils for three cases: the ripple from the TF coils alone (blue circles), the ripple from the TF coils and a dense dipole coil array (red squares), and the ripple from the TF coils and sparse dipole coils presented in \cref{subsec:single-stage-optimization} without the outboard midplane coil (green triangles). For all cases, the coil geometry is the same as the $N_\theta = 10$ configuration in \cref{fig:stage-II-pareto-front}. For the dipole coil cases, we perform a stage-II optimization to optimize the coil currents to minimize the field error on the tokamak surface, i.e. \cref{eq:stage-II-objective-function}. We compute $\delta$ at the outboard midplane for a sample tokamak configuration with a similar major and minor radius to HBT-EP, $\qty{1}{\meter}$ and $\qty{14}{\centi\meter}$, respectively.

With only TF coils, the ripple parameter reaches $\delta \approx 1\%$ at $20$ coils, matching the existing HBT-EP design outside of the minor differences introduced by variations in the coil/plasma geometry. With optimized dense or sparse dipole arrays, the required number of TF coils can be reduced to as few as $12$ coils for a dense array while remaining under the constraint of $\delta < 1 \%$. In these configurations, the dipole currents align in the poloidal direction, creating a TF-like magnetic field in between the TF coils to reduce the ripple; the maximum dipole coil current in any $\delta < 1 \%$ configuration is $\qty{27}{\kilo\ampere}$, approximately $15\%$ of the stellarator current requirements. This ripple-correction effect is possible for any $N_{\phi} = 2k \times N_{TF}$ with integer $k$, but $k=1$ is optimal and what we use here. For $k > 1$, the alternating current segments form TF-like structures at a higher toroidal periodicity, generating intermediate-scale ripple between the dipole coils that reduces the net correction.

The correction is reduced for the array without the outboard midplane, as expected because the TF coil ripple is strongest on the outboard side of the plasma. However, we still find that this array can correct ripple to reasonable limits down to $16$ total TF coils. The sparse array trades poloidal access (gained by removing outboard dipole coils) against toroidal access (gained by reducing the number of TF coils). Balancing these effects gives 16 TF coils and 32 toroidal dipole coils. Note that this matches the configuration used for the stellarator optimization in the previous sections. We also verified the on-axis $\delta$ for these $12$ (dense) and $16$-coil (sparse) minimums remains smaller than the TF-only, $20$-coil HBT-EP baseline, indicating our conclusions hold when evaluated over the plasma volume.

We have repeated this analysis for configurations at more reactor-relevant aspect ratios, and found that the dense array consistently allows a reduction of the number of TF coils by around 25\%. However, the correction of the sparse array does not necessarily support fewer TFs and is dependent on the size of the poloidal gap, requiring further study to identify the optimal configuration at lower aspect ratios. 

\subsection{Creating Shaped Tokamak Equilibria}
\label{subsec:shaped-tokamak-equilibria}
In this section, we show that toroidally aligned dipole currents can act as steady-state poloidal-field (PF) coils during tokamak operation. Specifically, when all coils at a single poloidal position are aligned, they create the magnetic field of two PF coils with currents in opposing directions at an offset equal to the poloidal width of the dipole coils in addition to a benign high toroidal mode number perturbation.

To demonstrate how this configuration can be used to create highly shaped tokamak equilibria, we run the Tokamaker code \cite{hansenTokaMakerOpensourceTimedependent2024} using the same $N_{\theta} = 10$ configuration used in previous sections to create both positive triangularity (PT) and negative triangularity (NT) equilibria. We use the same profiles as \cref{subsec:finite-beta-optimization} with the same $I_p = \qty{11.3}{\kilo\ampere}$ and $p_{axis} = \qty{306}{\pascal}$, with the same toroidal field as the stellarator equilibria, $B_{axis} = \qty{0.5}{\tesla}$. We set the plasma limiter to be $\qty{1}{\centi\meter}$ offset from the vacuum vessel. We model PF coils coincident with the toroidally parallel filaments of the dipole array, and assume that any contribution from fringing fields due to the central solenoid is negligible at the plasma. We also enforce that each pair of PF coils corresponding to a single dipole coil have equal and opposite currents. Because the dipole coils are HTS, their currents are treated as fixed during the pulse, so this analysis applies only to flattop. Additional copper PF coils will be needed during startup and rampdown, which we do not model here.

\begin{figure}[H]
    \centering
           \includegraphics[width=\textwidth]{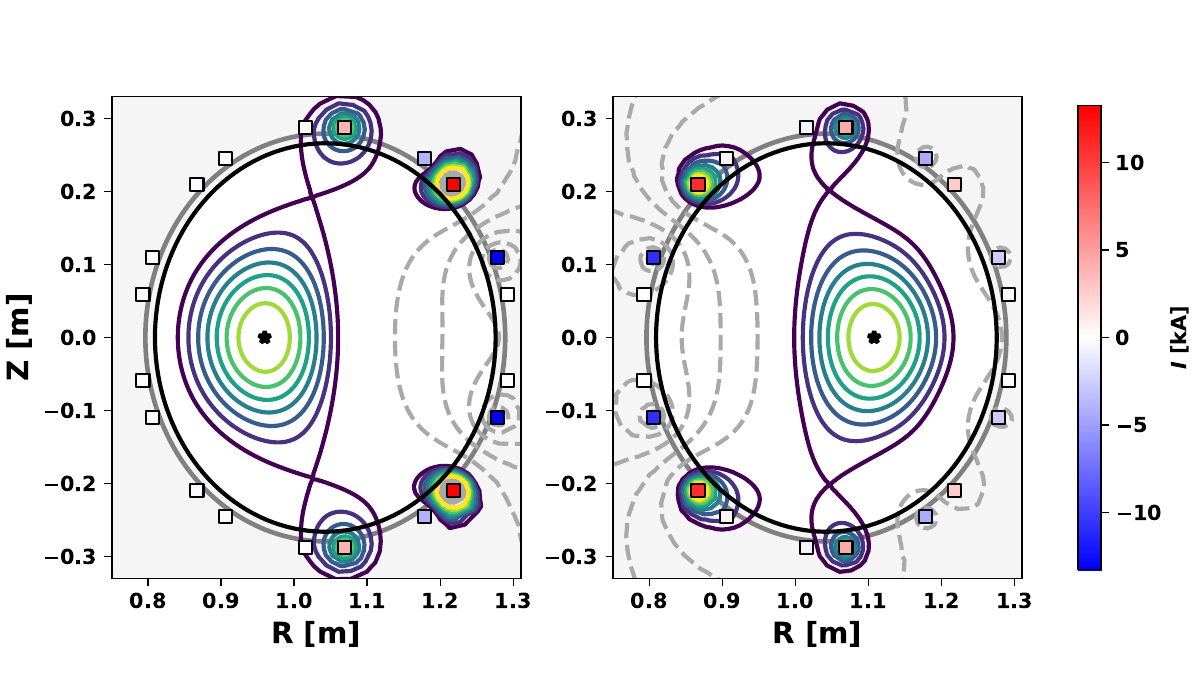}
    \caption{Demonstration of both a negative (left) and positive (right) triangularity tokamak equilibrium created using the axisymmetric dipole array approximated as PF coils with equal and opposite currents. The elongation $\kappa \approx 1.7$ in both equilibria, and $\delta \approx \pm 0.6$. The vacuum vessel is shown in grey, the limiter ($\qty{1}{\centi\meter}$ offset from the vacuum vessel) is in black, and the PF coils are shown as squares (size not to scale) colored by their current. We plot contours of the magnetic field, with those encompassing the plasma in color and those external to the plasma shown in dashed grey. The centroid of the plasma current is shown as a black star. The maximum coil currents are $\approx \qty{14}{\kilo\ampere}$ for both. \label{fig:shaped-tokamak-equilibria}}
   \end{figure}

\Cref{fig:shaped-tokamak-equilibria} shows the Grad-Shafranov solutions computed by Tokamaker for both NT and PT diverted double-null equilibria generated in this analysis. These results demonstrate that we can achieve both strong negative and positive triangularity at moderate elongation, with the solutions shown having an elongation $\kappa \approx 1.7$ and triangularity $\delta \approx \pm 0.6$. The maximum coil currents in these examples are $\qty{14.3}{\kilo\ampere}$ for NT and $\qty{13.3}{\kilo\ampere}$ for PT, less than $10\%$ of the maximum in the stellarator equilibria. Note that the dipole coil currents scale with the plasma current $I_p$. Mirroring the boundary targets about the major radius gives optimal solutions in which a PF coil acts as a divertor coil on the top and bottom of the vacuum vessel and a remaining subset of the coils acts as standard PF coils to shape the plasma and create a vertical field.

In the NT case, the last closed flux surface does not exhibit a poloidal
perturbation at the scale of the dipole coil spacing. The PT boundary, by
contrast, exhibits small poloidal oscillations in the field strength at that
scale. The mechanism is the same as in
\cref{subsec:single-stage-optimization}: as the equilibrium volume grows, the
plasma boundary expands toward the dipole coils, and the spatially concentrated
fields from individual coils produce poloidal ripple at the boundary. In the NT
configuration, the inboard dipole coils carry negligible currents and this
effect is small. In the PT case, however, the outboard boundary approaches the
PF coils supplying the vertical field, amplifying the ripple on that side. Whether this degrades confinement requires further analysis, but it will likely
determine the maximum achievable volume for a given set of elongation and
triangularity targets.

Furthermore, the outboard midplane PF coils carry near-zero current in both configurations,
a consequence of up-down symmetry. Because equal and opposite coil currents
produce an inherently up-down asymmetric field, they cannot contribute to
symmetric equilibria. This is consistent with the sparsity scheme proposed in
\cref{subsec:single-stage-optimization} and further motivates removing those
coils entirely. Up-down asymmetric equilibria, such as single-null diverted
configurations, could be accessed by allowing non-equal currents in the
poloidally mirrored coil pairs, analogous to standard PF coil operation in
tokamaks.

To assess the validity of the axisymmetric PF-coil approximation, we compute the full three-dimensional magnetic field produced by the discretized dipole and $16$ TF coils for both the PT and NT equilibria, following an analysis similar to \cref{subsec:finite-beta-optimization}: we compute the plasma normal field via the virtual casing principle and re-optimize the dipole coil currents using \cref{eq:finite-beta-objective-function}. We re-optimize the currents rather than directly imposing those from \cref{fig:shaped-tokamak-equilibria} so that the result captures both (i) the difference between the continuous PF-coil representation used by TokaMaker and the discrete dipole-coil representation, and (ii) the non-axisymmetric ripple introduced by the finite number of TF coils and its correction by the dipole array. Effect (i) alone increases the maximum individual dipole current needed to produce the same poloidal field at the boundary from $\qty{13}{\kilo\ampere}$ to $\qty{18}{\kilo\ampere}$ and effect (ii) superimposes a non-axisymmetric ripple-correction current on top of the axisymmetric shaping current. For the NT and PT configurations, respectively, we obtain surface-averaged, normalized field errors of $0.1\%$ and $0.3\%$, and maximum boundary ripple parameters $\delta$ of $0.5\%$ and $1.2\%$. The larger PT values arise because its boundary sits at a larger major radius, and therefore closer to the TF coils, than the reference geometry used in \cref{subsec:tf-ripple}. Repeating the PT calculation with an artificially large number of TF coils to remove this effect reduces the field error back to $0.1\%$. This finding confirms that the TF ripple, rather than the dipole discretization itself, is the dominant contribution to field error and motivates further study into combining TF-ripple correction with shaping currents in a single coil set. The NT ripple is within the $\delta \leq 1\%$ tolerance discussed in \cref{subsec:tf-ripple} and occurs near the divertor. It is therefore driven predominantly by the dipole currents, and its acceptably small value indicates that the axisymmetric PF-coil representation is a good approximation for the discrete coil set.

\section{Conclusion}
\label{sec:conclusion}

In this paper, we established the feasibility of a flexible, HBT-EP-inspired hybrid tokamak-stellarator experiment using an axisymmetric array of planar HTS dipole coils. Because the coil array has only a few geometric degrees of freedom, traditional two-stage optimization is insufficient; we instead used a single-stage approach that leverages the $\mathcal{O}(100)$ coil currents to obtain mutually consistent equilibria and coil sets. Our central finding is that the field-error and coil-current thresholds set a minimum and maximum coil-plasma distance, respectively, bounding the axisymmetric envelope available to the boundary, within which $\iota$, volume, current, and QS error trade off against one another. Within these constraints we obtained QA vacuum equilibria with low QS error and vacuum $\iota$ up to $0.15$ with a volume of $\qty{0.3}{\cubic\meter}$ (or $0.2$ with $\qty{0.25}{\cubic\meter}$), with peak pointwise forces of $\approx2.75\times10^5~\unit{\newton\per\meter}$, below estimated HTS limits.

We further showed that the same array operates at finite-$\beta$ with realistic profiles reaching on-axis $\iota\approx1$ and a vacuum transform in the regime relevant for stabilizing current-driven MHD modes, with the added profiles increasing the optimized maximum individual coil current and surface-averaged field error modestly ($\approx30\%$ and $0.1\%\to0.2\%$, respectively). In tokamak operation, the array can correct TF coil ripple, reducing the number of TF coils required to meet a target ripple parameter, and act as steady-state shaping coils to produce strongly shaped equilibria, including negative triangularity. Together, these results establish the device as a promising platform for hybrid tokamak-stellarator research spanning vacuum stellarators, current-carrying stellarators, and shaped tokamaks, motivating more detailed engineering analysis to constrain the operational space.

The flexibility that motivates this design also leaves a large parameter space largely unexplored, and we view the open physics questions it enables as the most compelling direction for future work. The ability to continuously vary the vacuum rotational transform provides a direct experimental handle on current-driven MHD stability, while the boundary-shaping flexibility across stellarator and tokamak equilibria offers a platform for studying how edge geometry influences exhaust and scrape-off-layer physics in both operating modes. We considered only QA equilibria here, as they are the most tokamak-like and the most relevant for hybrid operation. However, the peak forces we find reach at most $\approx70\%$ of the estimated HTS tolerance. This engineering headroom suggests that configurations demanding stronger non-axisymmetric shaping, such as quasi-helical or quasi-isodynamic equilibria, are within reach of a similar array, albeit likely at higher aspect ratio and larger coil currents.

Several extensions to the optimization and coil set also remain open. We fixed the coil geometry and varied only the currents; systematically scanning the vacuum-vessel ellipticity could improve the achievable transform and, by increasing the vessel circumference, accommodate additional poloidal dipole coils. We likewise held the TF coils fixed, whereas tilted TF coils have been used elsewhere to generate 3D fields \cite{morozLowaspectratioStellaratorsPlanar1997a, clarkProtoCIRCUSTiltedcoilTokamak2014, suzukiDesignSimpleStellarator2021a}. Combining tilted or shifted TF coils with the dipole array could expand the accessible equilibrium space or reduce shaping-coil currents, and we regard incorporating these degrees of freedom into the single-stage routine as a high-priority next step. Further open questions include the sensitivity of the optimization to initial conditions and a scan over field periodicity, which the $16$ TF and $32$ dipole coil layout naturally supports for $N_{fp}=1,2,4,\dots$.

Finally, several modeling choices warrant further refinement. Throughout, we modeled the coils as current-carrying filaments, which overestimates the spatial concentration of each coil's near-field. Incorporating a finite-build model into the single-stage optimization would smooth this contribution and could relax the minimum coil-plasma distance set by the poloidal-ripple constraint, potentially enlarging the achievable volume and rotational transform. Our engineering assessment also relied on the pointwise force $|d\mathbf{F}/dl|$ as a proxy, which might not be the true binding constraint for the device. More detailed modeling could include finite-element analysis to characterize net forces and torques on the coil support structure along with a full cost analysis, of which we expect the majority to be the cost of the HTS. For finite-$\beta$ operation, adding poloidal-field coils \cite{hennebergCompactStellaratortokamakHybrid2024a} or a finite-$\beta$ single-stage step could further reduce the coil currents and field error. Resolving these questions will determine how far the flexibility demonstrated here can be realized in a practical device.

%
% Each of the commands below will create an unnumbered section with the appropriate heading.
% Remove any sections that are not relevant for your article.
% All sections except suppdata will be removed if the [anonymous] option is used.
% See iopjournal-guidelines.pdf for more information.
%

\ack{J. H. would like to thank Antoine Baillod for teaching him much of what he knows about stellarator optimization and for many debugging sessions for this project. We also thank David Gates, Charles Swanson, and Thomas Kruger for input on dipole coil engineering feasibility and constraints and Sophia Henneberg for helpful discussions regarding hybrid equilibria and optimization.}

\funding{This material is based upon work supported by Columbia University, the U.S. Department of Energy, Office of Science, Office of Advanced Scientific Computing Research, Department of Energy Computational Science Graduate Fellowship under Award Number DE-SC0024386, and the Simons Foundation MPS Collaboration Program under Award Numbers 60651 and 11962. This research used resources of the National Energy Research Scientific Computing Center (NERSC), a Department of Energy Office of Science User Facility using NERSC award FES-ERCAP30322.}
% This section is a list of funder names and grant numbers

%\roles{Sample text inserted for demonstration.}
% List author names and the contributions made to the article, using terms from the NISO Contributor Roles Taxonomy (CRediT) https://credit.niso.org

%\data{Sample text inserted for demonstration.}
% For more information on IOP Publishing's research data policy see: https://publishingsupport.iopscience.iop.org/questions/research-data/

%\suppdata{Sample text inserted for demonstration.}

\printbibliography

\appendix
\crefalias{section}{appendix}

\section{Construction of the dipole coil array}
\label{sec:windowpane-array}
In this section, we expand upon the coil set description in \cref{sec:design-specification} and detail how we initialize the dipole coils on the axisymmetric vacuum vessel surface. We use the same SIMSOPT \cite{landremanSIMSOPTFlexibleFramework2021} coil representation as previous dipole coil work \cite{kaptanogluReactorscaleStellaratorsForce2025}, but with a fixed, specified geometry. 

The dipole coils are restricted to lie in a plane, and are represented by super-ellipses of poloidal width $a$ and toroidal height $b$ with rounded corners,
\begin{equation}
    r(\varphi) = \left(\left|\frac{\cos(\varphi)}{a}\right|^{n} + \left|\frac{\sin(\varphi)}{b}\right|^n\right)^{-1/n},
\end{equation}
where we have used $n=4$. The parameter $n$ can be adjusted to increase the curvature of the corners, where $n=2$ corresponds to a standard ellipse, and in the limit of $n\to\infty$ the curve becomes a rectangle of the same width and height. 
We set the curve order $M$ to be sufficiently large to accurately represent the curve ($\approx12$), and calculate $r_{c,m}$ and $r_{s,m}$ using the standard formula for Fourier coefficients, 
\begin{align}
    r_{c,0} &= \frac{1}{2\pi} \int_{0}^{2\pi} r(\varphi)d\varphi \\
    r_{c,m}& = \frac{1}{\pi} \int_{0}^{2\pi} r(\varphi)\cos(m\varphi) d\varphi \\
  r_{s,m} &= \frac{1}{\pi} \int_{0}^{2\pi} r(\varphi)\sin(m\varphi) d\varphi.
\end{align}
We then specify the coil translation via the center of the coil, $\textbf{r}_0 = (x_0, y_0, z_0)$, and rotation via a normalized quaternion 
\begin{equation}
    \textbf{q}=[a, b, c, d] = [\cos(\vartheta/2), u_x\sin(\vartheta/2), u_y\sin(\vartheta/2), u_z\sin(\vartheta/2)],
    \label{eq:quaternion}
\end{equation}
where $\vartheta$ is the counter-clockwise half angle rotation about a unit axis $\hat{\textbf{u}} = (u_x \hat{\textbf{e}}_x, u_y \hat{\textbf{e}}_y, u_z \hat{\textbf{e}}_z)$. After translation and rotation, the curve in Cartesian coordinates becomes
\begin{align}
    x(\varphi) &= (1-2(c^2+d^2))r(\varphi)\cos(\varphi) + 2(bc-ad)r(\varphi)\sin(\varphi) + x_0 \\
    y(\varphi) &= 2(ad+bc)r(\varphi)\cos(\varphi) + (1-2(b^2+d^2))r(\varphi)\sin(\varphi) + y_0 \\
    z(\varphi) &= 2(bd-ac)r(\varphi)\cos(\varphi) + 2(ab+cd)r(\varphi)\sin(\varphi) + z_0.
\end{align}

We place the coils locally tangent to the axisymmetric vacuum vessel, with a boundary $\mathbf{\Gamma}_{VV}$. We interpolate $\mathbf{\Gamma}_{VV}$ to compute the equally spaced coil center locations,
\begin{equation}
    \textbf{r}_0 = \mathbf{\Gamma}_{VV}(\theta_{i}, \phi_{j}),
\end{equation}
with $N_{\theta}$ and $N_{\phi}$ coils in the poloidal and toroidal directions, respectively. For each coil center location on the vessel, denoted by $(\theta_{i}, \phi_{j})$ with $i=1,...,N_{\theta}, j=1,...,N_\phi $, we calculate and set $a$ and $b$ to keep the spacing between coil filaments constant at $\qty{5}{\centi\meter}$. Note that this means the toroidal width $b$ varies substantially between the inboard and outboard midplanes.

The rotation quaternion is set such that the dipole coil unit normal vector is parallel to the interpolated vacuum vessel unit normal vector, 
\begin{equation}
\hat{\textbf{n}}_{VV,ij} = \frac{\frac{\partial\mathbf{\Gamma}_{VV}}{\partial \theta}\times\frac{\partial\mathbf{\Gamma}_{VV}}{\partial \phi}}{\left|\frac{\partial\mathbf{\Gamma}_{VV}}{\partial \theta}\times\frac{\partial\mathbf{\Gamma}_{VV}}{\partial \phi}\right|}\bigg|_{(\theta_{i}, \phi_{j})},
\end{equation}
and the axis with width $a_{ij}$ is aligned with the direction of $\partial\mathbf{\Gamma}_{VV}/\partial\theta$ at the coil center (i.e. the poloidal width aligns with the poloidal direction).
Given an initial coil unit normal, $\hat{\textbf{n}}_{C,ij}$, and assuming the initial $\hat{\textbf{e}}_x$ unit vector aligns with the axis of width $a_{ij}$, we first use \cref{eq:quaternion} to align the coil normal direction with the surface,
\begin{align}
    \vartheta_1 &= \arccos\left(\hat{\textbf{n}}_{C,ij} \cdot \hat{\textbf{n}}_{VV,ij}\right) \\
     \hat{\textbf{u}}_1 &= \frac{\hat{\textbf{n}}_{C,ij} \times \hat{\textbf{n}}_{VV,ij}}{\left|\hat{\textbf{n}}_{C,ij} \times \hat{\textbf{n}}_{VV,ij}\right|},
\end{align}
for non-parallel unit normals. Then, we obtain a second quaternion by aligning the rotated $\hat{\textbf{e}}_x$ from the first quaternion, $\hat{\textbf{e}}_x' = \textbf{q}_1 \hat{\textbf{e}}_x\textbf{q}_1^*$, with $\partial\mathbf{\Gamma}_{VV}/\partial\theta$, giving 
\begin{align}
    \vartheta_2 &= \arccos\left(\textbf{q}_1 \hat{\textbf{e}}_x\textbf{q}_1^* \cdot \left(\frac{\partial\mathbf{\Gamma}_{VV}/\partial\theta}{\left|\partial\mathbf{\Gamma}_{VV}/\partial\theta\right|}\right)_{(\theta_{i}, \phi_{j})}\right) \\
    \hat{\textbf{u}}_2& = \hat{\textbf{n}}_{VV}(\theta_{i}, \phi_{j}),
\end{align}
where $\textbf{q}_1^*$ is the conjugate quaternion to $\textbf{q}_1$. This specifies two rotation quaternions, and the total quaternion can then be determined via quaternion multiplication, $\textbf{q} = \textbf{q}_2 \textbf{q}_1$.

After initializing the coils, we apply stellarator symmetry and field period symmetry and fix the geometric degrees of freedom such that we only optimize the dipole coil currents of a half field period.

\section{Relation between transform and volume in a hybrid device}
\label{sec:near-axis-tradeoffs}

\begin{figure}[H]
 \centering
     \includegraphics[width=\textwidth]{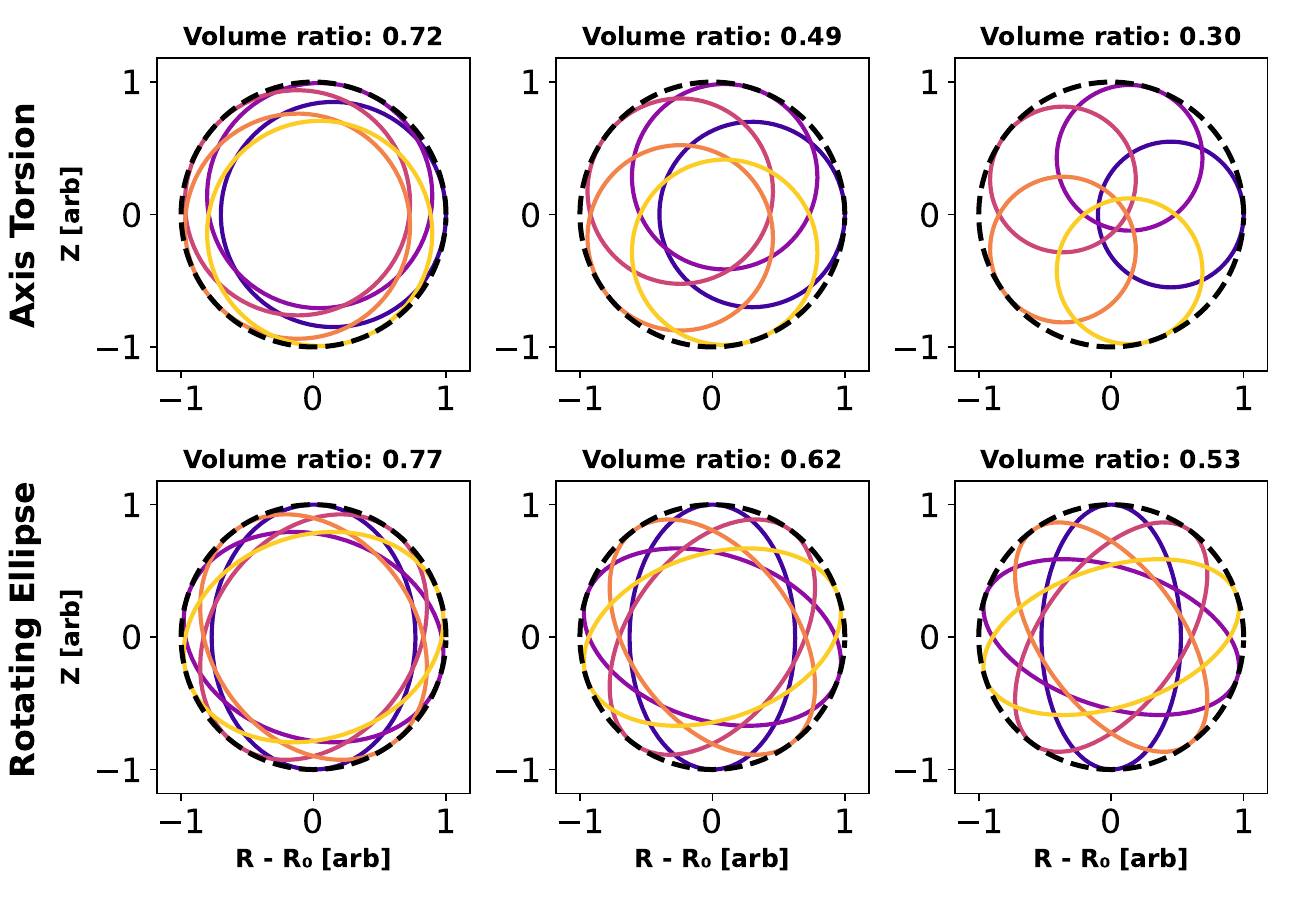}
 \caption{Analytic examples of axis torsion (top) and rotating ellipse (bottom) shaping cases with the amplitude, and therefore the axis rotational transform, increasing left to right. The dashed black curve indicates the enveloping axisymmetric boundary. For each panel, we plot the cross section at a few different toroidal locations in various colors. The panel titles show the corresponding volume ratio of the non-axisymmetric configuration compared to the axisymmetric envelope.}
 \label{fig:inverse-envelope-volume-ratio}
\end{figure}

In \cref{fig:inverse-envelope-volume-ratio}, we start with an axisymmetric, circular cross section with arbitrary major/minor radius $R_0$ and $a = 1$. We then apply perturbations, parametrized by an amplitude $A$, to model axis torsion and rotating ellipticity while enforcing the cross section to remain confined to the same axisymmetric envelope. For axis torsion, we add an axis displacement $A$:
\begin{align}
    R(\theta,\phi) - R_0 &= \left(a-A\right)\cos\theta + A\cos(\phi), \\
    Z(\theta,\phi)&=\left(a-A\right)\sin\theta + A\sin(\phi) \nonumber.
\end{align}
We increase $A$ from $0.15 \to 0.3 \to 0.45$ from left to right, so the minor radius decreases while the axis displacement increases. For the rotating ellipse, we modify the axis ratio at a fixed minor radius of $a$:
\begin{align}
    R(\theta,\phi) - R_0 &= a\left(\cos\theta\cos(\phi) - \frac{1}{1+A}\sin\theta\sin(\phi)\right), \\
    Z(\theta,\phi) &= a\left(\cos\theta\sin(\phi) + \frac{1}{1+A}\sin\theta\cos(\phi)\right) \nonumber,
\end{align}
where $A$ now controls the axis ratio of the ellipse via $b/a = 1+A$. We increase $A$ from $0.3 \to 0.6 \to 0.9$ from left to right. For each amplitude, we plot the cross section at $5$ equally spaced toroidal locations in various colors. We also compute the volume of the non-axisymmetric surface and display it in the panel titles normalized by the volume of the axisymmetric envelope.

It is clear that as the on-axis rotational transform increases, either via increasing the axis torsion or ellipticity of the magnetic surfaces, the volume of the stellarator must decrease to remain in an axisymmetric envelope. Note that rotational transform can also be generated by increasing the rate of rotation of the elliptical cross section via the field periodicity, which is not considered in this figure. However, in our case, this would likely require a corresponding increase in the number of dipole coils in the toroidal direction, which is not practical given the coupling between the number of TF coils and toroidal dipole coils.

\end{document}